%% file: chargedHiggs2L.tex
\documentclass[12pt]{article}
\usepackage{axodraw,amsmath,amssymb,amsfonts,color,graphicx,cite,color,feynarts,soul}
\input paperdef

\graphicspath{{figs/}}

\evensidemargin \oddsidemargin
\allowdisplaybreaks

\newlength{\captsize}          \let\captsize=\footnotesize
\newlength{\captwidth}         
\newlength{\beforetableskip}   
\newcommand{\capt}[1]{\begin{minipage}{\captwidth}
               \let\normalsize=\captsize
               \caption[#1]{#1}
               \end{minipage}\\  \vspace{\beforetableskip}}
\makeatletter
\newcommand{\fcaption}[1]{
         \refstepcounter{figure}
         \setbox\@tempboxa = \hbox{\footnotesize Figure~\thefigure. #1}
         \ifdim \wd\@tempboxa > 12cm
             {\footnotesize \textbf{Figure~\thefigure.} #1}
         \else
              {\begin{center}
              {\footnotesize \textbf{Figure~\thefigure.} #1}
               \end{center}}
         \fi}
\makeatother

\newcommand{\MHexp}{126}

\begin{document}
\thispagestyle{empty}

\def\thefootnote{\fnsymbol{footnote}}

\begin{flushright}
CERN--PH--TH/2013--114\\
DESY 13--100\\
MPP--2013--147 
\end{flushright}

\vspace{0.5cm}

\begin{center}

{\large\sc {\bf The Charged Higgs Boson Mass of the MSSM}}

\vspace{0.4cm}

{\large\sc {\bf in the Feynman-Diagrammatic Approach}}

\vspace{1cm}

{\sc
M.~Frank$^{1}$%
\footnote{email: m@rkusfrank.de}%
, L.~Galeta$^{2}$%
\footnote{email: leo@ifca.unican.es}%
, T.~Hahn$^{3}$%
\footnote{email: hahn@feynarts.de}%
, S.~Heinemeyer$^{2}$%
\footnote{email: Sven.Heinemeyer@cern.ch}%
, W.~Hollik$^{3}$%
\footnote{email: hollik@mppmu.mpg.de}%
,\\[.3em] H.~Rzehak$^{4}$%
\footnote{email: hrzehak@\,mail.cern.ch}%
\footnote{on leave from
Albert-Ludwigs-Universit\"at Freiburg, Physikalisches Institut, D--79104
Freiburg, Germany
}%
~and G.~Weiglein$^{5}$%
\footnote{email: Georg.Weiglein@desy.de}
}

\vspace*{.7cm}

{\sl
$^1$Institut f\"ur Theoretische Physik, Universit\"at Karlsruhe, \\
D--76128 Karlsruhe, Germany%
\footnote{former address}%

\vspace*{0.1cm}

$^2$Instituto de F\'isica de Cantabria (CSIC--UC), Santander,  Spain

\vspace*{0.1cm}

$^3$Max-Planck-Institut f\"ur Physik (Werner-Heisenberg-Institut),
F\"ohringer Ring 6, \\
D--80805 M\"unchen, Germany

\vspace*{0.1cm}

$^4$CERN, PH-TH, 1211 Geneva 23, Switzerland
\vspace*{0.1cm}

$^5$DESY, Notkestra\ss e 85, D--22607 Hamburg, Germany
}

\end{center}

\vspace*{0.1cm}

\begin{abstract}
\noindent
The interpretation of the Higgs signal at $\sim \MHexp \gev$ within
the Minimal Supersymmetric Standard Model (MSSM) depends crucially on
the predicted properties of the other Higgs states of the model, as
the mass of the charged Higgs boson, $\MHp$. This mass is calculated 
in the Feynman-diagrammatic approach within the MSSM with real parameters. 
The result includes the complete
one-loop contributions and the two-loop contributions of
\order{\alt\als}. 
The one-loop contributions lead to
sizable shifts in the $\MHp$ prediction, 
reaching up to $\sim 8 \gev$ for relatively small values of $\MA$. 
Even larger effects can occur
depending on the sign and size of the $\mu$~parameter that enters the
corrections affecting the relation between the bottom-quark mass and the
bottom Yukawa coupling.
The two-loop \order{\alt\als} terms can shift $\MHp$ by more than
$2 \gev$. The two-loop contributions amount to typically about 30\% of
the one-loop corrections for the examples that we have studied. These
effects can be relevant for precision analyses of
the charged MSSM Higgs boson. 
\end{abstract}

\def\thefootnote{\arabic{footnote}}
\setcounter{page}{0}
\setcounter{footnote}{0}

\newpage


\section{Introduction}

The ATLAS and CMS experiments at CERN have recently discovered a
new boson with a mass around 
$\MHexp \gev$~\cite{ATLASdiscovery,CMSdiscovery}. 
Within the presently still rather large experimental uncertainties
this new boson behaves like the 
Higgs boson of the Standard Model (SM)~\cite{Moriond13}. However,
the newly discovered 
particle can also be interpreted as the Higgs boson of extended models.
The Higgs sector of the Minimal Supersymmetric
Standard Model (MSSM)~\cite{mssm} with two scalar doublets
accommodates five physical Higgs bosons. In
lowest order these are the light and heavy $\cp$-even $h$
and $H$, the $\cp$-odd $A$, and the charged Higgs bosons $H^\pm$.
It was shown that the newly discovered boson can be interpreted 
in principle as the light, but also as the heavy
$\cp$-even Higgs boson of the MSSM, see, e.g., 
\citeres{Mh125,NMSSMLoopProcs,hifi,benchmark4,MH125other}.
In the latter case the charged Higgs boson must be rather light,
and the search for the charged Higgs boson could be crucial to
investigate this scenario~\cite{benchmark4}. In the former case the
charged Higgs boson is bound to be heavier than the top
quark~\cite{benchmark4}.
In both cases the discovery of a charged Higgs boson
would constitute an unambiguous sign of physics beyond the SM,
serving as a good motivation for searches for the charged Higgs boson.

The Higgs sector of the MSSM can be expressed at lowest
order in terms of the gauge couplings, the mass of the $\cp$-odd Higgs boson,
$\MA$, and $\tb \equiv v_2/v_1$,  
the ratio of the two vacuum expectation values. All other masses and
mixing angles can therefore be predicted, e.g.\ the charged Higgs boson
mass, 
\begin{align}
\label{MHptree}
\mHp^2 &= \MA^2 + \MW^2
\end{align}
at tree-level. $M_{Z,W}$ denote the masses of the $Z$~and $W$~boson,
respectively. Higher-order contributions can give 
large corrections to the tree-level relations, where the loop
corrected charged Higgs-boson mass is denoted as $\MHp$. 

Experimental searches for the neutral MSSM Higgs bosons have been
performed at LEP~\cite{LEPHiggsSM,LEPHiggsMSSM}, placing important
restrictions on the parameter space. At Run~II of the Tevatron the
search was continued, but is now superseeded by the LHC Higgs
searches.
Besides the discovery of a SM Higgs-like boson the LHC searches place
stringent bounds, in particular in the regions of small $\MA$ and large
$\tb$~\cite{CMSHiggsMSSM}. 
At a future linear collider (LC) a precise determination of the
Higgs boson properties 
(either of the light Higgs boson at $\sim \MHexp \gev$ or heavier MSSM
Higgs bosons within the kinematic reach) will be
possible~\cite{tesla,orangebook,acfarep,Snowmass05Higgs}. 
The interplay of the LHC and the LC in the neutral MSSM Higgs sector 
has been discussed in \citeres{lhcilc,eili}.

The charged Higgs bosons of the MSSM (or a more general Two Higgs
Doublet Model (THDM)) have also been searched for at
LEP~\cite{ALEPHchargedHiggs,DELPHIchargedHiggs,L3chargedHiggs,OPALchargedHiggs,LEPchargedHiggsPrel}, 
yielding a bound of
$\MHp \gsim 80 \gev$~\cite{LEPchargedHiggsProc,LEPchargedHiggs}.
The LHC places bounds on the charged Higgs mass, as for the neutral
heavy MSSM Higgs bosons, at relatively low values of its mass and at
large or very small $\tb$~\cite{CMSchargedHiggs,ATLASchargedHiggs}.
For $\mHp < \mt$ (with $\mt$ denoting the mass of the top quark)
the charged Higgs boson is mainly produced from top
quarks and decays mainly as $H^\pm \to \tau \nu_\tau$. 
For $\mHp > \mt$ the charged Higgs boson is mainly produced together with a
top quark and the dominant decay channels are $H^\pm \to tb, \tau \nu_\tau$, 
where the latter is the main search channel.
At the LC, for $\MHp \lsim \sqrt{s}/2$ a high-precision determination of the
charged Higgs boson properties will be
possible~\cite{tesla,orangebook,acfarep,Snowmass05Higgs}.

For the MSSM%
\footnote{We concentrate here on the case with real parameters. For
the case of complex parameters see
  \citeres{mhcMSSMlong,mhcMSSM2L} and references therein.}
the status of higher-order corrections to the masses and mixing angles
in the neutral Higgs sector is quite advanced. The complete one-loop
result within the MSSM is known~\cite{ERZ,mhiggsf1lA,mhiggsf1lB,mhiggsf1lC}.
The by far dominant one-loop contribution is the \order{\alt} term due
to top and stop loops ($\alt \equiv h_t^2 / (4 \pi)$, $h_t$ being the
top-quark Yukawa coupling). The computation of the two-loop corrections
has meanwhile reached a stage where all the presumably dominant
contributions are 
available~\cite{mhiggsletter,mhiggslong,mhiggslle,mhiggsFD2,bse,mhiggsEP0,mhiggsEP1,mhiggsEP1b,mhiggsEP2,mhiggsEP3,mhiggsEP3b,mhiggsEP4,mhiggsEP4b,mhiggsRG1,mhiggsRG1a}.
In particular, the \order{\alt\als}, \order{\alt^2}, \order{\alb\als},
\order{\alt\alb} and \order{\alb^2} contributions to the self-energies
are known for vanishing external momenta.  For the (s)bottom
corrections, which are mainly relevant for large values of $\tb$,
an all-order resummation of the $\tb$-enhanced term of
\order{\alb(\als\tb)^n} is
performed~\cite{deltamb1,deltamb2,deltamb2b}. 
The remaining theoretical uncertainty on the lightest $\cp$-even Higgs
boson mass has been estimated to be about
$\sim 3 \gev$~\cite{mhiggsAEC,PomssmRep,mhiggsWN}. 
The public codes \fh~\cite{feynhiggs,mhiggslong,mhiggsAEC,mhcMSSMlong} 
(including all of the above corrections) and {\tt CPsuperH}~\cite{cpsh}
exist. 
A full two-loop effective potential calculation %
(including even the momentum dependence for the leading
pieces and the leading three-loop corrections) has been
published~\cite{mhiggsEP5}. However, no computer code 
is publicly available. Most recently another leading three-loop
calculation, depending on the various SUSY mass hierarchies, became
available~\cite{mhiggsFD3l}, resulting in the code {\tt H3m} (which
adds the three-loop corrections to the {\tt FeynHiggs} result).

Also the mass of the charged Higgs boson is affected by higher-order
corrections.  
However, the status is somewhat less advanced as compared to the neutral
Higgs bosons. 
First, in \citere{mhp1lA} leading corrections to the relation given in 
\refeq{MHptree} have been evaluated. 
The one-loop corrections from $t/b$ and $\Stop/\Sbot$~loops have been
derived in \citeres{mhp1lB,mhp1lC}. A nearly complete one-loop
calculation, neglecting the terms suppressed by higher powers of the
SUSY mass scale, was presented in \citere{mhp1lD}. The
first full one-loop calculation in the Feynman-diagrammatic (FD)
approach has been performed in \citere{mhp1lE}, and re-evaluated more
recently in \citeres{markusPhD,mhcMSSMlong}. 
At the two-loop level,
within the FD approach, the leading \order{\alt \als} two-loop contributions 
for the three neutral Higgs bosons in   
the case of complex soft SUSY-breaking parameters have 
been obtained~\cite{mhcMSSM2L}. Because of the ($\cp$-violating) mixing
between all three neutral Higgs bosons, in the MSSM with
complex parameters usually the charged Higgs mass is chosen as
independent (on-shell) input parameter, which by construction does not 
receive any higher-order corrections. The calculation however 
involves the evaluation 
of the \order{\alt \als} contributions to the charged $H^\pm$ self energy.
In the $\cp$-conserving case, on the other hand, where usually 
$\MA$ instead of $\MHp$ is chosen as independent input parameter, 
the corresponding self-energy contribution can be utilized to obtain
corrections of \order{\alt \als} to the mass $\MHp$. 

In the present paper we combine the two-loop terms of \order{\alt \als}
with the complete one-loop contribution of \citere{mhcMSSMlong}
to obtain an improved prediction for the mass 
of the charged Higgs boson. 
The results are incorporated in the code \fh\ (current version: 2.9.4). 
An overview of the calculation is given in Sect.\ 2,  whereas
in Sect.\ 3 and \refse{sec:numanal}
we discuss the size and relevance of the one- and two-loop corrections
and investigate the
impact of the various sectors of the MSSM on the 
prediction for $\MHp$. 
Our conclusions are given in  \refse{sec:conclusions}.


\section{Higher-order contributions for \boldmath{$\MHp$}}
\label{sec:mhp}

\subsection{From tree-level to higher-orders}
\label{sec:tree}

In the MSSM (with real parameters) one conventionally chooses the mass
of the $\cp$-odd Higgs boson, $\MA$, and $\tb$ ($\equiv v_2/v_1$, see
\refeq{eq:higgsdoublets}) as independent input 
parameters. Thus the mass of the charged Higgs boson can be predicted in
terms of the other parameters and receives a shift from the higher-order
contributions.

The two Higgs doublets of the MSSM are decomposed in the following way,
\begin{align}
\label{eq:higgsdoublets}
\cHe = \begin{pmatrix} H_{11} \\ H_{12} \end{pmatrix} &=
\begin{pmatrix} v_1 + \tfrac{1}{\sqrt{2}} (\phi_1-i \chi_1) \\
  -\phi^-_1 \end{pmatrix}, \notag \\ 
\cHz = \begin{pmatrix} H_{21} \\ H_{22} \end{pmatrix} &= 
\begin{pmatrix} \phi^+_2 \\ v_2 + \tfrac{1}{\sqrt{2}} (\phi_2+i
  \chi_2) \end{pmatrix} ,
\end{align}
with the two vacuum expectation values $v_1$ and $v_2$.
The hermitian 2$\times$2-matrix of the charged states $\phi_{1,2}^\pm$, 
$\matr{M}_{\phi^\pm\phi^\pm}$, contains the following elements, 
\begin{align}
\label{eq:massen_phipm}
\matr{M}_{\phi^\pm\phi^\pm} &= 
\begin{pmatrix}
m_1^2 + \tfrac{1}{4} g_1^2 (v_1^2 - v_2^2) + \tfrac{1}{4} g_2^2 (v_1^2
+ v_2^2) & 
-  m_{12}^2 - \tfrac{1}{2} g_2^2 v_1 v_2 \\[.5em]
-  m_{12}^2 - \tfrac{1}{2} g_2^2 v_1 v_2 &
m_2^2 + \tfrac{1}{4} g_1^2 (v_2^2 - v_1^2) + \tfrac{1}{4} g_2^2 (v_1^2 + v_2^2)
\end{pmatrix}.
\end{align}
$m_1$, $m_2$, $m_{12}$ denote the soft SUSY-breaking parameters in the
Higgs sector, and $g_2$, $g_1$ are the $SU(2)$ and $U(1)$ gauge
couplings, respectively.
The mass eigenstates in lowest order in the charged sector follow from 
unitary transformations on the original fields, 
\begin{align}
\label{eq:RotateToMassES}
\begin{pmatrix} H^\pm \\ G^\pm \end{pmatrix} = 
\begin{pmatrix} -\sinb & \cosb \\ \cosb & \sinb \end{pmatrix}
\cdot
\begin{pmatrix} \phi_1^\pm \\ \phi_2^\pm \end{pmatrix} .
\end{align}
This yields the (square of the) mass eigenvalue for the charged Higgs
boson, $\mHp^2$, as given by \refeq{MHptree}. 
Quantum corrections
substantially modify the tree-level mass. 
The charged Higgs-boson pole mass, $\MHpm$,
including higher-order contributions entering via 
the renormalized charged Higgs-boson
self-energy, $\hSi_{H^+H^-}$,
is obtained by solving the equation
\begin{align}
p^2 - \mHpm^2 + \hSi_{H^+H^-}(p^2) \; = \; 0~.
\label{MHp}
\end{align}
This yields
$\MHp^2$ as the real part of the complex zero of \refeq{MHp}. 
The renormalized charged Higgs-boson
self-energy, $\hSi_{H^+H^-}$, is composed of the 
unrenormalized self-energy, $\Si_{H^+H^-}$, and counterterm contributions
as specified below. 
In perturbation theory, the self-energy is expanded 
as follows
\begin{align}
\Si(p^2)  &= \Si^{(1)}(p^2)  + \Si^{(2)}(p^2)  + \ldots~,
\end{align}
in terms of the $i$th-order contributions
$\Si^{(i)}$, and analogously for the renormalized quantities.
Details for
the one-loop self-energies are given below in \refse{sec:1l}, and 
for the two-loop contributions in \refse{sec:2l}. 

A possible mixing with the charged Goldstone boson would contribute
to the prediction for the charged Higgs-boson mass from
two-loop order onwards via terms of the form 
$\KL \hSi^{(1)}_{H^\pm G^\mp}(p^2) \KR^2$.
The mixing contributions with $G^\pm$ yield a two-loop
contribution that is subleading compared to the leading terms at 
\order{\alt \als} that we take into account, as described in
\refse{sec:2l}. Consequently, we neglect those two-loop Higgs--Goldstone
mixing contributions
throughout our analysis.


\subsection{One-loop corrections}
\label{sec:1l}

Here we review the calculation of the full one-loop
corrections to $\MHp$, following \citeres{mhcMSSMlong,markusPhD}. 
All self-energies and 
renormalization constants are understood to be one-loop quantities,
dropping the order index.
Renormalized self-energies, $\hSi(p^2)$,
can be expressed in terms of the corresponding
unrenormalized self-energies, $\Si(p^2)$, the field
renormalization constants, and the mass counterterms.
For the charged Higgs-boson self-energy 
entering \refeq{MHp}
this expression reads
\begin{align}
\ser{H^+ H^-}(p^2)  &= \se{H^+ H^-}(p^2) + \dZ{H^+ H^-} (p^2 -
\mHpm^2) - \dmHpmsq~.
\label{SErenHpm}
\end{align}
The independent mass parameters are renormalized according to
\begin{align}
\MA^2  &\to \MA^2 + \de\MA^2~, \non \\
\MW^2  &\to \MW^2 + \de\MW^2~,
\end{align}
while the mass counterterm for the charged Higgs boson,
arising from $\mHp^2 \to \mHp^2 + \de\mHp^2$, is a dependent quantity.
It is given in terms of the counterterms for $\MA$ and $\MW$ by
\begin{align}
\de\mHp^2 &= \de\MA^2 + \de\MW^2~.
\end{align}
We renormalize the $W$~boson and the $\cp$-odd Higgs boson masses
on-shell,
yielding the mass counterterms 
\begin{align}
\label{eq:mass_osdefinition}
  \dMWsq = \re \se{WW}(\MW^2),
  \quad \dMAsq = \re \se{AA}(\MA^2)~,
\end{align}
where $\se{WW}$ is the transverse part of the $W$~boson self-energy.

For field renormalization, required for
finite self-energies at arbitrary values of the external momentum $p^2$,
we assign one field-renormalization constant
for each Higgs doublet,
\begin{align}
\label{eq:HiggsDublettFeldren}
  \cHe \rightarrow (1 + \tfrac{1}{2} \dZ{\cHe}) \cHe, \quad
  \cHz \rightarrow (1 + \tfrac{1}{2} \dZ{\cHz}) \cHz.
\end{align}
For the charged Higgs field this implies
\begin{align}
H^\pm \to (1 + \tfrac{1}{2} \dZ{H^+H^-}) \; H^\pm~,
\end{align}
with 
\begin{align}
\dZ{H^+H^-} &= \sinbsq \dZ{\cHe} + \cosbsq \dZ{\cHz}~.
\end{align}
For the determination of the field renormalization constants
we adopt the $\DRbar$~scheme, 
\begin{align}
  \dZ{\cHe} &= \dZ{\cHe}^{\DRbar}
       \; = \; - \KKL \re \Sip_{HH \; |\al = 0} \KKR^{\rm div}, \non \\[.5em]
  \dZ{\cHz} &= \dZ{\cHz}^{\DRbar} 
       \; = \; - \KKL \re \Sip_{hh \; |\al = 0} \KKR^{\rm div},
\label{eq:deltaZHiggs}
\end{align}
i.e.\ the renormalization constants 
consist of divergent parts only, see the discussion in 
\citere{mhcMSSMlong}.
$\Sip_{\phi\phi \; |\al = 0}$ ($\phi = h,H$) denotes the derivative of the
unrenormalized self-energies of the neutral $\cp$-even Higgs bosons, with
the mixing angle $\al$ set to zero. 
As default value of the renormalization scale we have chosen 
$\mu^{\DRbar} = \mt$.

For the
self-energies as specified in \refeq{SErenHpm} we have evaluated 
the complete 
one-loop contributions with the help of the programs
\fa~\cite{feynarts} and  \fc~\cite{formcalc}.
As regularization scheme we have used constrained
differential regularization~\cite{cdr}, which has been shown to be
equivalent to
dimensional reduction~\cite{dred} at the \onel\ level~\cite{formcalc},
thus preserving supersymmetry~\cite{dredDS,dredDS2}.
The corresponding Feynman-diagrams for the
charged Higgs boson (and similarly for the $W$~boson, 
where additional diagrams with gauge boson and ghost loops
contribute) are shown in 
\reffi{fig:fdSEc}. The diagrams for the neutral Higgs bosons, entering
$\de\MA^2$ and $\dZ{\cHe}$, $\dZ{\cHz}$ (i.e.\ the neutral Higgs boson
self-energies) , are depicted in \reffi{fig:fdSEn}.

\setlength{\unitlength}{0.093mm}
\begin{figure}[htb!]
\begin{center}
\input{fdSEc}
\end{center}
\fcaption{Generic Feynman diagrams for the $H^\pm$ self-energy
($l$ = \{$e$, $\mu$, $\tau$\}, 
 $d$ = \{$d$, $s$, $b$\},
 $u$ = \{$u$, $c$, $t$\}).
Similar diagrams for the $W$~boson self-energy are obtained by
replacing the external Higgs boson by a $W$~boson; not all
  combinations of particle insertions exist.}
\label{fig:fdSEc}
\end{figure}

\setlength{\unitlength}{0.099mm}
\begin{figure}[htb!]
\begin{center}
\input{fdSEn}
\end{center}
\fcaption{Generic Feynman diagrams for the $h$, $H$, $A$, self-energies 
($f$ = \{$e$, $\mu$, $\tau$, $d$, $s$, $b$, $u$, $c$, $t$\} ).
Not all combinations of particle insertions exist for all neutral Higgs
bosons. 

\vspace*{1cm}
}
\label{fig:fdSEn}
\end{figure}


\subsection{Two-loop corrections}
\label{sec:2l}

We now turn to the \order{\alt\als} corrections at the two-loop
level. 
Again, we drop the order-index for 
all Higgs boson and SM gauge boson self-energies and renormalization
constants, which are in this section understood to be of
two-loop order.
The \order{\alt\als} terms are obtained in the limit of vanishing
gauge couplings and neglecting the dependence on the external
momentum~\cite{mhiggslong}, keeping only terms $\propto h_t^2 \als$,
with the top Yukawa coupling $h_t$ as defined above. 
We neglect the
bottom Yukawa coupling in the two-loop 
Higgs-boson self-energies.
In this approximation, the 
counterterm for $\MA$ is determined as follows
\begin{align}
\de\MA^{2} = \Si_{AA}(0)~,
\end{align}
while the renormalization constants $\de\MW^2$ and $\dZ{H^+H^-}$ do not
contribute,
\begin{align}
\de\MW^{2} &= 0,\qquad  \dZ{H^+H^-} = 0~.
\end{align}
Consequently, 
the two-loop contribution to the renormalized $H^\pm$
self-energy can be written in the following way,
\begin{align}
\hSi_{H^+H^-}(0) &= \Si_{H^+H^-}(0) - \de\mHp^{2}
\qquad {\rm with} \qquad \de\mHp^{2} = \de\MA^{2}~.
\end{align}
From \refeq{MHp} we get the two-loop correction to the charged
Higgs-boson mass, 
\begin{align}
 \De\mHp^{2{\rm ,2-loop}} \, =  \Si_{AA}(0) - \Si_{H^+H^-}(0)
\end{align}
with the self-energies evaluated at the two-loop level.

\begin{figure}[htb!]
\begin{center}
\includegraphics[width=0.9\linewidth]{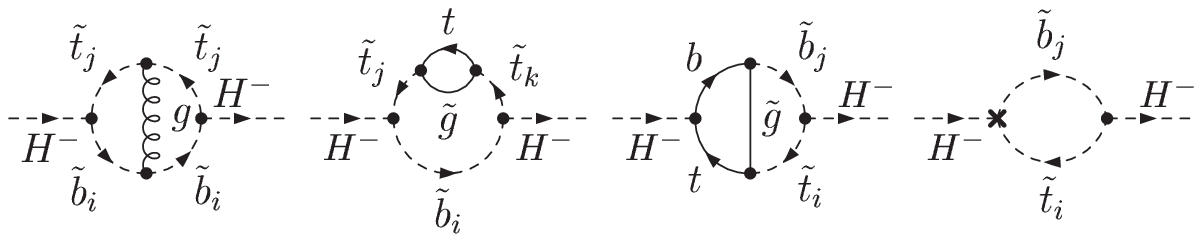}
\end{center}
\fcaption{Examples of generic two-loop diagrams and diagrams with
  counterterm insertion for the charged Higgs-boson self-energy
($i,j,k = 1,2$).}
\label{fig:fd_Hpm}
\end{figure}

\begin{figure}[htb!]
\begin{center}
\includegraphics[width=0.9\linewidth]{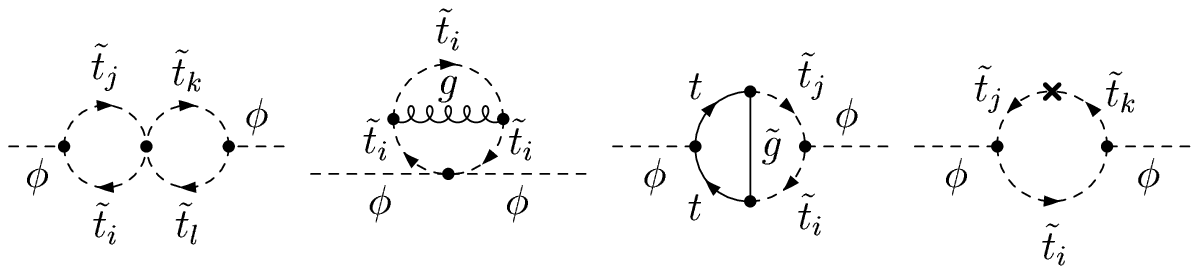}
\end{center}
\fcaption{Examples of generic two-loop diagrams and diagrams with counterterm
insertion for the $A$~boson self-energy ($\;i,j,k,l = 1,2$).}
\label{fig:fd_A}
\end{figure}

We thus have to evaluate the
\order{\alt\als} contributions to the $H^\pm$ and $A$ self-energies. 
Examples of generic Feynman diagrams for the $H^\pm$ self-energy are depicted in
\reffi{fig:fd_Hpm}, and in \reffi{fig:fd_A} for the
$A$~boson self-energy. 
These contributions have been evaluated using the packages
\fa~\cite{feynarts} and \tc~\cite{twocalc}.

\subsection{Subloop renormalization in the scalar top/bottom sector}

Besides the computation of the genuine two-loop diagrams 
at \order{\alt\als}
for the self-energies, one-loop renormalization 
is required for the $\Stop$ and $\Sbot$~sector
providing the counterterms for one-loop
subrenormalization. This yields additional diagrams with 
counterterm insertions;
examples are the fourth diagrams in
\reffis{fig:fd_Hpm}, \ref{fig:fd_A}. 
The bilinear part of the $\Stop$ and $\Sbot$ Lagrangian,
\begin{align}
\cL_{\Stop/\Sbot\text{ mass}} &= - \begin{pmatrix}
{{\tilde{t}}_{L}}^{\dagger}, {{\tilde{t}}_{R}}^{\dagger} \end{pmatrix}
\matr{M}_{\tilde{t}}\begin{pmatrix}{\tilde{t}}_{L}\\{\tilde{t}}_{R}
\end{pmatrix} - \begin{pmatrix} {{\tilde{b}}_{L}}^{\dagger},
{{\tilde{b}}_{R}}^{\dagger} \end{pmatrix}
\matr{M}_{\tilde{b}}\begin{pmatrix}{\tilde{b}}_{L}\\{\tilde{b}}_{R} 
\end{pmatrix}~,
\end{align}
contains the stop and sbottom mass matrices
$\matr{M}_{\tilde{t}}$ and $\matr{M}_{\tilde{b}}$,
given by 
\begin{align}\label{Sfermionmassenmatrix}
\matr{M}_{\tilde{q}} &= \begin{pmatrix}  
 M_L^2 + m_q^2  & 
 m_q \Xq \\[.2em]
 m_q \Xq &
 M_{\tilde{q}_R}^2 + m_q^2  
\end{pmatrix}~, \\ 
{\rm with} &\mbox{} \non \\
\Xq &= \Aq - \mu \kappa~, \qquad \kappa = \{\cot\beta, \tan\beta\} 
        \quad {\rm for} \quad q = \{t,b\}~.
\end{align}
Here $M_L^2$, $M_{\tilde{q}_R}^2$ are soft-breaking parameters, where
$M_L^2$ is the same for 
$\matr{M}_{\tilde{t}}$ and $\matr{M}_{\tilde{b}}$ (see below), and
$A_q$ is the trilinear soft-breaking parameter.
The D-terms do not contribute to \order{\alt\als} and therefore
have to be neglected in the calculation of the stop mass values 
entering the contribution of this order~\cite{mhcMSSM2L}.
The mass matrix can be diagonalized with the help of a unitary
transformation ${\matr{U}}_{\tilde{q}}$, which can be parametrized
by a mixing angle $ {\theta}_{\tilde{q}}$, 
\begin{align}\label{transformationkompl}
\matr{D}_{\tilde{q}} &= 
\matr{U}_{\tilde{q}}\, \matr{M}_{\tilde{q}} \, 
{\matr{U}}_{\tilde{q}}^\dagger = 
\begin{pmatrix} \msqe^2 & 0 \\ 0 & \msqz^2 \end{pmatrix}\,,\quad\
{\matr{U}}_{\tilde{q}}= 
\begin{pmatrix} U_{\tilde{q}_{11}}  & U_{\tilde{q}_{12}} \\  
                U_{\tilde{q}_{21}} & U_{\tilde{q}_{22}}  \end{pmatrix}
= \begin{pmatrix} 
  \cos {\theta}_{\tilde{q}} & 
  \sin {\theta}_{\tilde{q}} \\ 
  - \sin {\theta}_{\tilde{q}} & 
  \cos {\theta}_{\tilde{q}} \end{pmatrix}~. 
\end{align}

We follow here the renormalization prescription used
in \citeres{hr,mhiggsFDalbals}.
In the MSSM the $t/\Stop$~sector is described in terms of four real
parameters (where we assume that $\mu$ and $\tb$ are defined via other
sectors): the real soft SUSY-breaking
parameters $M_L^2$ and $M_{{\tilde{t}}_R}^2$, the trilinear coupling $\At$,
and the top Yukawa coupling $h_t$.
Instead of the quantities
$M_L^2$, $M_{{\tilde{t}}_R}^2$ and $h_t$, in the on-shell scheme
applied in this paper we choose the
on-shell squark masses $\mste^2$, $\mstz^2$ and the top-quark mass
$\mt$ as independent parameters.
It should furthermore be noted that the counterterms are evaluated
at \order{\als}, such as to yield the desired \order{\alt\als}
contributions when combined with the one-loop diagrams with
counterterm insertion.

\smallskip
The following renormalization conditions are imposed:
\begin{itemize}
\item[(i)] The top-quark mass is defined on-shell, yielding the 
 mass  counterterm $\de \mt$,
\renewcommand{\widetilde}[1]{#1}
\begin{align}\label{dmt}
\de  \mt = \frac{1}{2} \mt \bigl(
 \widetilde{\text{Re}}{\Si}_t^L (\mt^2) 
+ \widetilde{\text{Re}} {\Si}_t^R (\mt^2) 
+ 2 \widetilde{\text{Re}}{\Si}_t^S (\mt^2)\bigr) ~,
\end{align}
referring to the Lorentz decomposition of the self energy ${\Si}_{t}$
\begin{align}
{\Si}_{t} (k) &= \not\! k {\omega}_{-}
{\Si}_t^L (k^2) +\not\! k {\omega}_{+}
{\Si}_t^R (k^2) + \mt {\Si}_t^S (k^2) 
\label{decomposition}
\end{align}
into a left-handed, a right-handed, and a
scalar 
part, ${\Si}_t^L$, ${\Si}_t^R$, ${\Si}_t^S$,
respectively.  
\item[(ii)]
The stop masses are also determined via on-shell
conditions~\cite{mhiggslong,hr}, yielding  
\begin{align}\label{dmst}
\de  m_{\tilde{t}_i}^2 &= 
\widetilde{\text{Re}}\Si_{\tilde{t}_{ii}}(m_{{\tilde{t}}_{i}}^2)
\quad\ \text{with} \quad\ i = 1,\,2~.
\end{align}
\item[(iii)]
The third condition affects the trilinear coupling $\At$.
Rewriting the squark mass matrix in terms of the mass eigenvalues and
the mixing angle 
using \refeq{transformationkompl},
\begin{align}
\label{CTexpansion}
{\matr{M}}_{\Stop} &= \begin{pmatrix} 
  \cos^2\tst \mste^2 + \sin^2\tst \mstz^2 & 
  \sin\tst \cos\tst(\mste^2 - \mstz^2) \\[.3em]
  \sin\tst \cos \tst(\mste^2 - \mstz^2) & 
  \sin^2\tst \mste^2 + \cos^2\tst \mstz^2 
\end{pmatrix} \, ,
\end{align}
yields the counterterm matrix $\delta {\matr{M}}_{\Stop}$ 
by introducing counterterms $\delta\mste^2, \delta\mstz^2$
for the masses and $\delta\tst$ for the angle.
One obtains the counterterm for the off-diagonal contribution in the
stop sector,
\begin{align}
\label{Ydef}
 (m^2_{\Stop_1} - m^2_{\Stop_2}) \, \de \theta_{\Stop} =
 [\matr{U}_{\Stop}\, \de\matr{M}_{\Stop}\, \matr{U}_{\Stop}^\dagger]_{12}
 \equiv \de Y_{\Stop} \, ,
\end{align}
for which the following renormalization condition has been 
used~\cite{mhiggsFDalbals,hr}:
\begin{align}
\label{Yrenormalization}
\de Y_{\Stop} = \frac{1}{2}\, [  
 {\text{Re}}{\Si}_{\Stop_{12}}(\mste^2)  +
 {\text{Re}}{\Si}_{\Stop_{12}}(\mstz^2)] \, .
\end{align}
Finally we derive the relation between the counterterms
$\de\At$ and $\de\tst$.
The two counterterms 
are mutually related via \refeq{Sfermionmassenmatrix} 
and \refeq{CTexpansion}. The off-diagonal entries of the corresponding
counterterm matrices yield
\begin{align}
(\At - \mu \cot\beta)\, \de \mt + \mt\, \de \At &=
\sin\tst \cos\tst
(\de \mste^2 - \de \mstz^2) +
\KL \cos^2\tst - \sin^2\tst \KR \de Y_t.
\end{align}
As a result, we obtain for $\de\At$
\begin{align}
\label{dAt}
\de \At &=  \frac{1}{\mt}
\KKL \frac{1}{2} \sin 2\tst (\de \mste^2 - \de \mstz^2) 
+ \cos 2\tst \, \de Y_t
- \frac{1}{2\mt} \sin 2\tst (\mste^2 - \mstz^2) \de \mt \KKR .
\end{align}

\end{itemize}

In the $b/\Sbot$ sector, we also encounter 
four real parameters (with $\mu$ and $\tb$ defined via other sectors):
the soft-breaking mass  
parameters $M_L^2$ and $M_{{\tilde{b}}_R}^2$, 
the trilinear coupling $A_b$, and the bottom 
Yukawa coupling $h_b$ or the $b$-quark mass, respectively 
(which is neglected for the set
of two-loop corrections presented in this paper).
SU$(2)$ invariance requires the ``left-handed'' soft-breaking 
parameters in the stop and the sbottom sector to be
identical (denoted as $M_L^2$).
With the approximations described above this yields, e.g., 
$\msbl = M_L$.
In the evaluation of the \order{\alt\als} contributions to
the Higgs-boson self-energies, the counterterms of the sbottom sector 
appear only in the self-energy of the charged Higgs boson. 
In our approximation for the two-loop contributions, where the 
$b$-quark mass is neglected, 
$\SbotL$ and $\SbotR$ do not mix, 
and $\SbotR$ decouples and does not contribute.
The two-loop contribution to the
charged Higgs-boson self-energy thus depends only on a single
parameter of the sbottom sector, which can be chosen as the squark mass
$m_{\tilde{b}_L}$. 
By means of SU$(2)$ invariance, the corresponding mass  counterterm
is already determined:
\begin{align}
\label{dmsb}
\de \msbl^2 = 
 \cos^2\tst \, \de \mste^2 + \sin^2\tst \, \de \mstz^2
 - \sin 2\tst \, \de Y_t - 2 \mt \, \de \mt~.
\end{align} 

With the set of renormalization constants determined in \refeqs{dmt},
(\ref{dmst}), (\ref{dAt}) and (\ref{dmsb})
the counterterms arising from the one-loop subrenormalization of the stop and
sbottom sectors are fully specified.

\medskip
Finally, at \order{\alt\als} gluinos appear as
virtual particles at the two-loop level; hence, no renormalization in
the gluino sector is
needed. The corresponding 
soft-breaking gluino mass parameter is denoted $M_3$. In the case of
real MSSM parameters considered here the gluino
mass is given as $\mgl = M_3$.


\subsection{Higher-order corrections in the \boldmath{$b/\Sbot$} sector}
\label{sec:db}

We furthermore include in our prediction for $\MHp$ 
corrections beyond the one-loop level originating from the
bottom/sbottom sector contributions to $\se{AA}$ and $\se{H^+H^-}$.
Potentially large
higher-order effects proportional to $\tb$ can arise in the 
relation between the bottom-quark mass and the bottom Yukawa coupling
as described in \citeres{deltamb2,deltamb2b}. The leading $\tb$-enhanced
contribution in the limit of heavy SUSY masses can be expressed in terms of a
quantity $\db$ and resummed to all orders using an effective Lagrangian
approach.
The relevant part of the effective Lagrangian is given by
\begin{align}
\cL = \frac{g}{2\MW} \frac{\mbms}{1 + \db} \Bigg[ 
 \tb\; A \, i \, \bar b \ga_5 b 
   + \wz \, V_{tb} \, \tb \; H^+ \bar{t}_L b_R 
    \Bigg] + {\rm h.c.}~.
\label{effL}
\end{align}
Here 
\begin{align}
\label{mbdrbar}
\mbms^{\DRbar, {\rm SM}}(Q) &= 
\mbms^{\MSbar, {\rm SM}}(Q) \KL 1 - \frac{\als}{3\,\pi} \KR~,\\
\mbms  &=  
\mbms^{\DRbar, {\rm SM}}(Q = \mt) 
 \; \KL 1 + \edz \KL \Si^L_{b,{\rm fin}}(\mb) + \Si^R_{b,{\rm fin}}(\mb)
    \KR\KR~.
\end{align}
$\mbms^{\DRbar, {\rm SM}}(Q)$ denotes a running bottom quark mass at
the scale~$Q$ in the \DRbar\ scheme that incorporates SM QCD corrections 
(i.e., no SUSY QCD effects are included in the running).
The corresponding mass in the \MSbar\ scheme is denoted by 
$\mbms^{\MSbar, {\rm SM}}(Q)$.
$\Si^L_{b,{\rm fin}}(\mb)$ and $\Si^R_{b,{\rm fin}}(\mb)$ are the
finite parts of the self-energies defined in analogy to
\refeq{decomposition}. 
$V_{tb}$ denotes the $(3,3)$~element of the CKM matrix.
In the numerical evaluations performed with the program
{\tt FeynHiggs} below we use 
$\mbms^{\DRbar, {\rm SM}}(Q = \mt) \approx 2.68 \gev$. 

The leading $\tb$-enhanced one-loop contribution
in the limit of heavy SUSY masses takes the simple
form~\cite{deltamb1}
\begin{align}
\db &= \frac{2\als}{3\,\pi} \, \mgl \, \mu \, \tb \,
                    \times \, I(\msbe, \msbz, \mgl) +
      \frac{\alt}{4\,\pi} \, \At \, \mu \, \tb \,
                    \times \, I(\mste, \mstz, \mu) \,+\, \ldots~,
\label{def:Deb}
\end{align}
where $\als$ is evaluated at the scale $\sqrt{\msbe\,\msbz}$, and the
function $I$ is given by 
\begin{align}
I(a, b, c) &= \ed{(a^2 - b^2)(b^2 - c^2)(a^2 - c^2)} \,
               \KL a^2 b^2 \log\frac{a^2}{b^2} +
                   b^2 c^2 \log\frac{b^2}{c^2} +
                   c^2 a^2 \log\frac{c^2}{a^2} \KR \\
 &\sim \ed{\mbox{max}(a^2, b^2, c^2)} ~. \non
\end{align}
The ellipses in \refeq{def:Deb} denote subleading terms that we
take over from \citere{deltab4}.
Expanded up to one-loop order, the effective mass $\mbms/(1 + \db)$ is
close to the \DRbar\ mass (including SUSY contributions in the running), 
see \citeres{mhiggsEP4,mhiggsFDalbals}.
A recent two-loop calculation of $\db$ can be found in \citere{db2L}.


\section{Approximation for the two-loop corrections}
\label{sec:approx}

In \refse{sec:mhp} we have described the approximations 
for getting the two-loop \order{\alt\als} terms, which can be
written as terms proportional to $\mt^4$.
It is well known that for the neutral
Higgs bosons this procedure indeed yields the dominant part of the
one-loop~\cite{ERZ,mhiggsf1lA,mhiggsf1lB,mhiggsf1lC} and the two-loop
corrections~\cite{mhiggsletter,mhiggslong}. 

For the charged Higgs boson mass, $\MHp$, the described procedure 
provides the analogous contribution to the mass shift as well, 
\begin{align}
\De\MHp^2 \sim \frac{\mt^4}{v^2}
 \sim  \frac{\mt^4}{\MW^2}~.
\label{eq:mt4MW2}
\end{align}
There are, however, other contributions of similar structure at the
one-loop level~\cite{mhp1lA,mhp1lB,mhp1lC,mhp1lD,mhp1lE},
\begin{align}
\De\MHp^2 \sim \frac{\mt^2\,\mb^2}{\MW^2} \mbox{~~~~or~~~~}
\De\MHp^2 \sim \frac{\mt^4}{\MW^2} \frac{\MW^2}{\msusy^2} \mbox{~~~~or~~~~}
\De\MHp^2 \sim \frac{\mt^4}{\MW^2} \frac{\MA^2}{\msusy^2}~,
\label{leading1lold}
\end{align}
which are not covered by our approximations for the two-loop terms
because they would correspond to $\mb \neq 0$ (first), 
non-vanishing gauge-couplings (second),
and $p^2\neq 0$ in the $A$ self-energy (third term). 
This is justified for large scalar-quark mass scales $\msusy$
where the second and third type of terms are suppressed.
On the other hand, the term (\ref{eq:mt4MW2})
extracted by our approximation
can in general be large also for large $\msusy$, both at the one-loop 
and the two-loop level, as we will explain below.


\subsection{The one-loop case}

Applying the approximations outlined in \refse{sec:2l} at the one-loop level
yields the counter\-terms (all quantities in this section are
  understood to be one-loop quantities),
\begin{align}
\de\MW^2 &= 0,\qquad \de\MA^2 = \Si_{AA}(0), \qquad 
 \dZ{H^+H^-} = 0~,
\end{align}
and thus
\begin{align}
\hSi_{H^+H^-} &= \Si_{H^+H^-}(0) - \de\mHp^{2}
 \qquad {\rm with} \qquad 
\de\mHp^{2} = \de\MA^{2}~. 
\end{align}
From \refeq{MHp} we get the one-loop corrected value of the charged
Higgs-boson mass, 
\begin{align}
\MHp^2 &= \mHp^2 + \De\mHp^2~,
\end{align}
with
\begin{align}
 \De\mHp^2 \, =  \Si_{AA}(0) - \Si_{H^+H^-}(0)~.
\end{align}
In the following we use the factor $c$ to simplify the notation
($v^2 = v_1^2 + v_2^2$), 
\begin{align}
c &=  - \frac{3\, \mt^2}{16\, \pi^2\, v^2\, \TQb} =
  - \frac{3\, e^2\, \mt^2}{32\, \pi^2\, \sw^2\, \MW^2\, \TQb}~.
\label{prefactor}
\end{align}
From the third (s)quark generation, with $m_b=0$, one obtains the explicit
expressions 
\begin{align}
\Si_{AA}(0) &= c \;  
\Big\{ 2 A_0(\mt) - A_0(\mstz) - A_0(\mste)
      - (\At + \mu \tb)^2\frac{A_0(\mstz) - A_0(\mste)}{\mstz^2 - \mste^2} 
\Big\},\non \\
\Si_{H^+H^-}(0) &= c \;
\Big\{ 2 A_0(\mt) - A_0(\msbe) - \stt^2 A_0(\mste) - \ctt^2 A_0(\mstz) \non\\
&\qquad - \KL \ctt \mt + \stt (\At + \mu \tb) \KR^2
         \frac{A_0(\mste) - A_0(\msbe)}{\mste^2 - \msbe^2} \non \\
&\qquad - \KL \stt \mt - \ctt (\At + \mu \tb) \KR^2
         \frac{A_0(\mstz) - A_0(\msbe)}{\mstz^2 - \msbe^2}
\Big\}.
\end{align}
Here we use the abbreviation 
$\stt \equiv \sin\tst, \ctt \equiv \cos\tst$,
and the one-loop integral function $A_0(m)$ is defined as
in \citere{denner}. 
In the approximation of $m_b = 0$ and neglected gauge couplings 
the mass of the left-handed sbottom is given by
\renewcommand{\msbe}{m_{{\tilde b}_L}}
\begin{align}
\msbe^2 &= \ctt^2 \mste^2 + \stt^2 \mstz^2 - \mt^2 \; (= M_L^2)~.
\end{align}
Using these relations results then
in the following expression for $\De\mHp^2$:
\begin{align}
\De\mHp^2 = - c \Bigl\{& 
\msbe^2 \Bigl[1 + \frac{[(\At + \mu  \tb)\stt + \mt \ctt ]^2}
       {\msbe^2 - \mste^2} 
 + \frac{[\mt \stt - \ctt (\At + \mu
\tb)]^2}{\msbe^2 - \mstz^2}\Bigr] 
\log\Bigl(\frac{\msbe^2}{\mste^2}\Bigr) 
      \non \\& 
+ \mstz^2 \Bigl[-\stt^2 + \frac{(\At + \mu \tb)^2}{\mste^2 -
    \mstz^2} 
- \frac{[\mt \stt - \ctt (\At + \mu \tb)]^2}{\msbe^2 -
      \mstz^2}\Bigr] \log\Bigl(\frac{\mstz^2}{\mste^2}\Bigr)\Bigr\} .
\end{align}
It is possible to eliminate the dependence on $\At$ and $\tst$ from the
expression of the charged Higgs-boson mass correction, 
\begin{align}
\De\mHp^2 &= \frac{c \mt^2}{(\msbe^2 - \mste^2)(\msbe^2 - \mstz^2)} \;
             \frac{\mu^2}{\SQb\CQb} \times \non \\
     &\quad   \KKL \msbe^2 \log \KL \frac{\msbe^2}{\mste\mstz} \KR
                 - \frac{  \mste^2 (\msbe^2 - \mstz^2) 
                         + \mstz^2 (\msbe^2 - \mste^2)}
                        {\mste^2 - \mstz^2} 
                  \log \KL \frac{\mste}{\mstz} \KR \KKR~.
\end{align}
This shows explicitly the $\mt^4$ dependence of this contribution as
well as the overall factor $\mu^2/\CQb$, which strongly determines the
phenomenology of the \order{\alt} charged Higgs-boson mass corrections.
In the following, we specify the analytic result, assuming a common SUSY
mass scale 
$ M_L = M_{\tilde{t}_R} \, =: \msusy $. 
With this simplification one obtains
\begin{align}
\mste^2 = \msusy^2 + \mt^2 - \mt |\Xt|, \quad
\mstz^2 = \msusy^2 + \mt^2 + \mt |\Xt|, \quad
\msbe^2 = \msusy^2~,
\label{MStopMSbot}
\end{align}
(and $\stt^2 = \ctt^2 = 1/2$ in this case). This yields

\begin{align}
\De\mHp^2 &= \frac{c \mt^2 \, \mu^2}{\SQb\CQb} \times 
             \ed{2 \mt^2 |\Xt| \, (\Xt^2 - \mt^2)} \times \non \\
&\quad \Big[ \mt (\msusy^2 + \mt^2 - \Xt^2) \, 
       \log \KL \frac{\msusy^2 + \mt (\mt - |\Xt|)}
                     {\msusy^2 + \mt (\mt + |\Xt|)} \KR \non \\
&\quad - \msusy^2 |\Xt| \, 
       \log \KL \frac{\msusy^4}
                     {\msusy^4 + 2 \msusy^2 \mt^2 + \mt^4 - \mt^2 \Xt^2} \KR
       \Big]~.
 \end{align}

Expanding in inverse powers of $\msusy$ and inserting the prefactor
$c$ from \refeq{prefactor}
we find
\begin{align}
\De\mHp^2  
&\approx  - \frac{3\, e^2 \mu^2}{64\, \pi^2\, \sw^2\, \sin^4\be} \;
\frac{\mt^4}{\MW^2} \; \Bigl[
\frac{1}{\msusy^2} - \frac{2 \mt^2}{3 \msusy^4} + \frac{\mt^2 (3 \mt^2 +
  \Xt^2)}{6 \msusy^6} \Bigr]~. 
\label{maxmixfinal}
\end{align}
Thus one obtains the term proportional to $\mt^4/\MW^2$.
In the special case of $\Xt = 0$ and restricting to the leading 
term in the expansion in inverse powers of $\msusy$ 
(vanishing stop mixing) this reduces to
\begin{align}
\De\mHp^2 
&\approx - \frac{3\,
  e^2\,\mu^2}{64\, \pi^2\, \sw^2\,  \sin^4 \be} \,
\frac{\mt^4}{\MW^2\, \mst^2}~,
\label{nomixfinal}
\end{align}
where $\mst^2 \equiv \msusy^2 + \mt^2$.
If $|\mu| \approx \mst$
this term is not suppressed by large SUSY mass scales.


\subsection{The two-loop case}
\label{sec:mt4MW2-2L}

The derivation of \refeqs{maxmixfinal} and (\ref{nomixfinal}) shows that
besides the $\mt^2$ in the prefactor arising from the Yukawa couplings, the
second factor $\sim \mt^2$ stems from the stop mass matrix. In other
words, it is induced by the $SU(2)$ breaking in the MSSM quark and
squark sector. Thus, the derived term $\sim \mt^4/\MW^2$
is related to the 
mass difference between top and bottom squarks resulting from
$\mt/\mb \gg 1$. The diagrams playing the leading role here are the
second and sixth Feynman diagram in \reffis{fig:fdSEc}, \ref{fig:fdSEn}.

\refeqs{maxmixfinal} and (\ref{nomixfinal}) indicate which parameter
combinations of $\At$, $\mu$ and $\tb$ can give rise to a sizable
\order{\alt} contribution to $\MHp^2$ and possibly constitute a large
part of the full one-loop corrections.
For the corresponding parameter ranges it can be expected that
also the new two-loop corrections of \order{\alt\als} are sizable
and should be taken into account. 

At the two-loop level the $\sim \mt^4$ contributions are augmented by the
corresponding term with a renormalized $\mt$ parameter, leading to
$\sim 4 \mt^3 \de\mt$. 
The source of these corrections is still related to 
the $SU(2)$ breaking inducing the mass difference for
scalar tops and bottoms, which enters 
the two-loop level Higgs-boson self-energies
through mass-counterterm insertions, as illustrated in
the fourth diagram in \reffi{fig:fd_A}.
The inserted one-loop counterterms 
are given by \refeq{dmst} for top-squarks
and by \refeq{dmsb} for bottom-squarks.
They differ essentially 
by a term $2\mt\de\mt$, 
which induces an effective
mass splitting between the scalar top and bottom sector at the
counterterm level. The full contribution $\sim \de\mt$ can be obtained
by renormalizing $\mt$ in \refeqs{maxmixfinal} and (\ref{nomixfinal}), or
by an explicit extraction of this term. 
In the case of vanishing stop mixing, corresponding to
\refeq{nomixfinal}, we have checked that both calculations indeed
agree. In this case they yield (keeping in mind the prefactor
$c \propto \mt^2$ in \refeq{prefactor})
\begin{align}
\De\mHp^{2,{\rm 2-loop,\de\mt}} &\sim \frac{(\At + \mu\tb)^2}{\mt^2} 
\KKL \de\mst^2 \; \frac{\mt^2}{\mst^2}
    -(\de\mst^2 - 2\mt\de\mt)
     \log \KL \frac{\mst^2}{\mst^2 - \mt^2} \KR \KKR~, \non \\
\label{nomixfinal2l}
                       &= \frac{\mu^2}{\SQb\CQb} 
\KKL \frac{\de\mst^2}{\mst^2}
    -\frac{\de\mst^2 - 2\mt\de\mt}{\mt^2} 
     \log \KL \frac{\mst^2}{\mst^2 - \mt^2} \KR \KKR~, \\
&\approx \frac{\mu^2}{2 \SQb\CQb} 
\KKL \frac{4\mt\de\mt}{\mst^2} \KKR~. \non
\end{align}
For the case of non-vanishing stop mixing, see \refeq{maxmixfinal}, we
find accordingly 
\begin{align}
\De\mHp^{2,{\rm 2-loop,\de\mt}} &\sim  \frac{\mu^2}{\SQb\CQb}
      \Bigg[ \frac{2 \mt}{\msusy^2} - \frac{2 \mt^3}{\msusy^4}
           + \frac{\mt^3 \KL 4 \mt^2 + \Xt^2 \KR}{2 \msusy^6} \Bigg]
      \times \de\mt~.
\label{maxmixfinal2l}
\end{align}

In conclusion, although the two-loop corrections to $\MHp^2$ 
covered by our approach 
are only part of the complete two-loop Yukawa corrections, 
they constitute a finite well-defined subset that 
can induce non-negligible mass shifts for the $H^\pm$ boson.
Numerical examples will be given in \refse{sec:twoloop}.


\section{Numerical analysis}
\label{sec:numanal}

Our results obtained in this paper extend the known results in the
literature in various ways. While the one-loop result in \citere{mhp1lE}
was complete, the numerical evaluation focused on particular
parameter values, mostly excluded nowadays by the LEP Higgs
searches~\cite{LEPHiggsSM,LEPHiggsMSSM,LEPchargedHiggsPrel,LEPchargedHiggsProc}.
\citere{mhpAB} focused on the mass splitting $\MHp - \MA$ induced by
$\db$~effects. 
We perform a more general numerical analysis, including the full
one-loop corrections. Furthermore for the first time explicit two-loop
corrections to $\MHp$ are analyzed.
The higher-order
corrected Higgs-boson sector has been evaluated with the help of the
Fortran code \fh~\cite{mhiggslong,feynhiggs,mhiggsAEC,mhcMSSMlong}
(current version: 2.9.4).

The goal for the precision in predicting $\MHp$ in the MSSM should
be the prospective experimental resolution or better. 
For the LHC no dedicated study has been performed recently.
Older evaluations indicate that a precision $\lsim 5\%$ might be
possible in the region of large $\tb$~\cite{MHpLHClarge}.
Other studies, focusing on the $\tau \nu_\tau$ decay mode yielded a
precision at the 1--2\% level~\cite{MHpLHCILCnewer}.
At the LC for $\MHp < \mt$ a precision of $\sim 1 \gev$ could be
possible~\cite{MHpILCsmall}, while for $\MHp > \mt$ 
(but $\MHp < \sqrt{s}/2$) a $\sim 1.5\%$ precision might be
reachable using the $t\bar b$ decay mode~\cite{Snowmass05Higgs}.
The $\tau \nu_\tau$ decay mode, on the other hand, could yield a precision of 
$\sim 0.5 \%$~\cite{MHpLHCILCnewer}.

\medskip
Due to the large number of MSSM parameters, certain benchmark 
scenarios~\cite{benchmark2,benchmark3,benchmark4} 
(for real parameters)
have been used for the interpretation of MSSM Higgs
boson searches at LEP, the Tevatron and the LHC. 
Since at tree level the Higgs sector of the MSSM is 
governed by two parameters (in
addition to the gauge couplings), the definition of the
benchmarks is usually such that the two tree-level parameters, $\MA$ and
$\tb$, are varied while the values of all other parameters are fixed at
certain benchmark settings. 
The most commonly used benchmark scenario for the
$\cp$-conserving MSSM has been the $\mhmax$
scenario~\cite{benchmark2,benchmark3,benchmark4}, and we therefore
employ this scenario in our analysis.
While the interpretation of the newly discovered
Higgs-like state as the light MSSM Higgs boson is compatible with the
$\mhmax$ scenario only within a strip at relatively low $\tb$,
it should be noted that changing the stop mixing
parameter $\Xt$ from the ``maximal'' value of $\Xt/\msusy \sim 2$
to slightly smaller values
(with the other parameters fixed) yields $\Mh \sim \MHexp \gev$ over
large parts of the parameter space, see \citere{benchmark4}. 
Consequently, this scenario is expected to provide a good
indication of the possible size of
the radiative corrections to $\MHp$.
The scenario is defined as follows:

\noindent
\underline{The $\mhmax$ scenario:}\\[.2em]
In this scenario the parameters are chosen such that the mass of the
light $\cp$-even Higgs boson acquires its maximum possible values 
as a function of $\tb$ (for fixed $M_{\rm SUSY}$, $\mt$ 
and $\MA$ set to its maximum value, $\MA = 1$~TeV).
This was used in particular to obtain conservative $\tb$ 
exclusion bounds~\cite{tbexcl} at LEP~\cite{LEPHiggsMSSM}.
The parameters are (including the most recent value for
$\mt$~\cite{mt1732}):
\begin{eqnarray}
&& \mt = \mu^{\DRbar} = 173.2 \gev, \quad \msusy = 1 \tev, \quad
\mu = 200 \gev, \quad M_2 = 200 \gev, \nonumber \\
\label{mhmax}
&& \Xt = 2\, \msusy,  \quad
   \Ab = \At, \quad m_{\tilde g} = 0.8\,\msusy~.
\end{eqnarray}
$\msusy$ ($\equiv M_L = M_{\tilde q_R}$)
denotes the diagonal soft SUSY-breaking parameters in the
$\Stop/\Sbot$ mass matrices, see \refeq{Sfermionmassenmatrix}, that
are all chosen to be equal. 
$\msusy$ and $\Xt$ in this scenario correspond to the
parameters used to express the
$\mt^4/\MW^2$ corrections as given in \refeqs{maxmixfinal},
(\ref{maxmixfinal2l}). 
In order to avoid conflicts with the LHC searches for squarks of the
first and second generation, contrary to the original
definition~\cite{benchmark2}, 
$\msusy$ should only be considered to fix the soft SUSY-breaking
parameters for the squarks of the third generation, while the first two
generations play a small role for the MSSM Higgs phenomenology.
To fix a value for the squarks of the first two generations, 
for sake of simplicity, we kept the value of
$\msusy$, but choosing higher values has a minor impact (see below).
The gluino mass parameter, $\mgl$, might also be in conflict with recent
LHC SUSY searches. However, since also the impact of $\mgl$ is
relatively small, we keep its value at the original definition.
(A slightly higher value is chosen in the updated version of this
scenario in \citere{benchmark4}.)

As discussed above, there are also potentially large
corrections in the $b/\Sbot$ sector, depending on the 
value and sign of the parameter $\mu$~\cite{benchmark3}. Consequently,
besides analyzing the $\MHp$ dependence on $\MA$ and $\tb$, we also
study the effect of a variation of $\mu$, 
allowing both an enhancement and a suppression of the bottom Yukawa
coupling.
Concerning the $\mhmax$ benchmark scenario, 
as discussed in \citeres{cmsHiggs,benchmark3} (see also
\citeres{earlier}), the $\db$ effects are 
particularly pronounced, since the two terms in
\refeq{def:Deb} are of similar size. 

The other MSSM parameters that are not specified above, such as the
slepton masses, have only a minor
impact on MSSM Higgs-boson phenomenology. In our numerical analysis
below we fix them such that all soft SUSY-breaking parameters in the
diagonal entries of the slepton mass matrices are set to $\msusy$, and
the trilinear couplings for all sfermions are set to $\At$, if not
indicated differently for $\Ab$ ($= \Atau$).

\bigskip
For the analysis of the size of the two-loop corrections we employ
in addition also a scenario that yields particularly interesting
phenomenology for the 
charged Higgs boson. In this scenario the {\em heavy} $\cp$-even Higgs
boson is interpreted as the newly discovered particle at $\sim \MHexp \gev$,
see, e.g., \citeres{Mh125,NMSSMLoopProcs,hifi,benchmark4,MH125other}.
The starting point for this scenario is the ``best-fit'' value obtained
in a seven parameter fit in the MSSM, 
where the interpretation of the signal at $\sim \MHexp \gev$ as the heavy
$\cp$-even Higgs boson of the MSSM has been confronted with the measured
signal strengths, taking into account also constraints from electroweak
precision observables and flavour physics~\cite{hifi}.
The parameters are (close to the parameters
in the ``low-$\MH$ scenario defined in \citere{benchmark4}):

\noindent
\underline{The light heavy-Higgs scenario:}\\[.2em]
\begin{align}
\mt &= 173.2 \gev~, \non \\
\msqd &:= \MstL  (= \MsbL) = \MstR = \MsbR = 670 \gev~, \non \\
\msld &:= \MstauL (= \MsneutL) = \MstauR = 323 \gev~, \non \\
\Af &= 1668 \gev~, \non \\
\MA &= 124.2 \gev~, \non \\
\tb &= 9.8~, \non \\
\mu &= 2120 \gev~, \non \\
M_2 &= 304 \gev~,\non \\[.5em]
\MsqL &= \MsqR~(q = c, s, u, d) \; = \; 1000 \gev~, \non \\
\MslL &= \MslR~(l = \mu, \nu_\mu, e, \nu_e) \; = \; 300 \gev~, \non \\
\mgl &= 1000 \gev~, \non \\
M_1 &= \frac{5}{3} \frac{\sw^2}{\cw^2} M_2 \approx \frac{1}{2}  \MTwo~,
\label{lhH}
\end{align}
where the latter four were fixed in the fit.
$\msqd$ denotes the diagonal soft SUSY-breaking parameter for the third
generation squarks, $\MsqL$ and $\MsqR$ for the first and second
generation squarks, $\msld$ for the third generation sleptons, and 
$\MslL$ and $\MslR$ for the first and second generation sleptons.
$\Af$ denotes the trilinear Higgs-sfermion coupling which is taken to be
equal for all sfermions.


\subsection{One-loop corrections}
\label{sec:oneloop}

\begin{figure}[htb!]
\centerline{\includegraphics[width=0.5\linewidth]{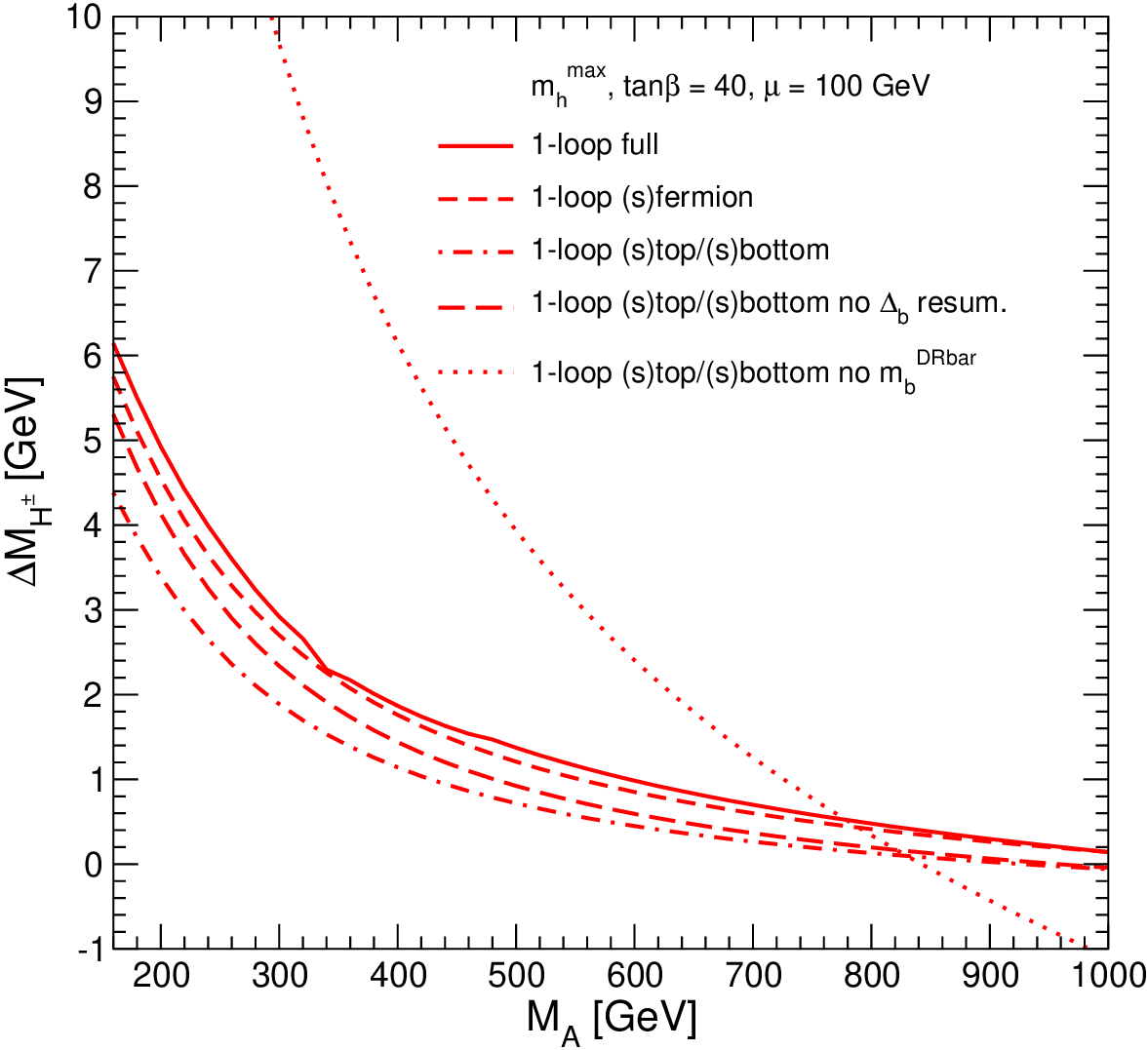}
            \hspace{1em}
            \includegraphics[width=0.5\linewidth]{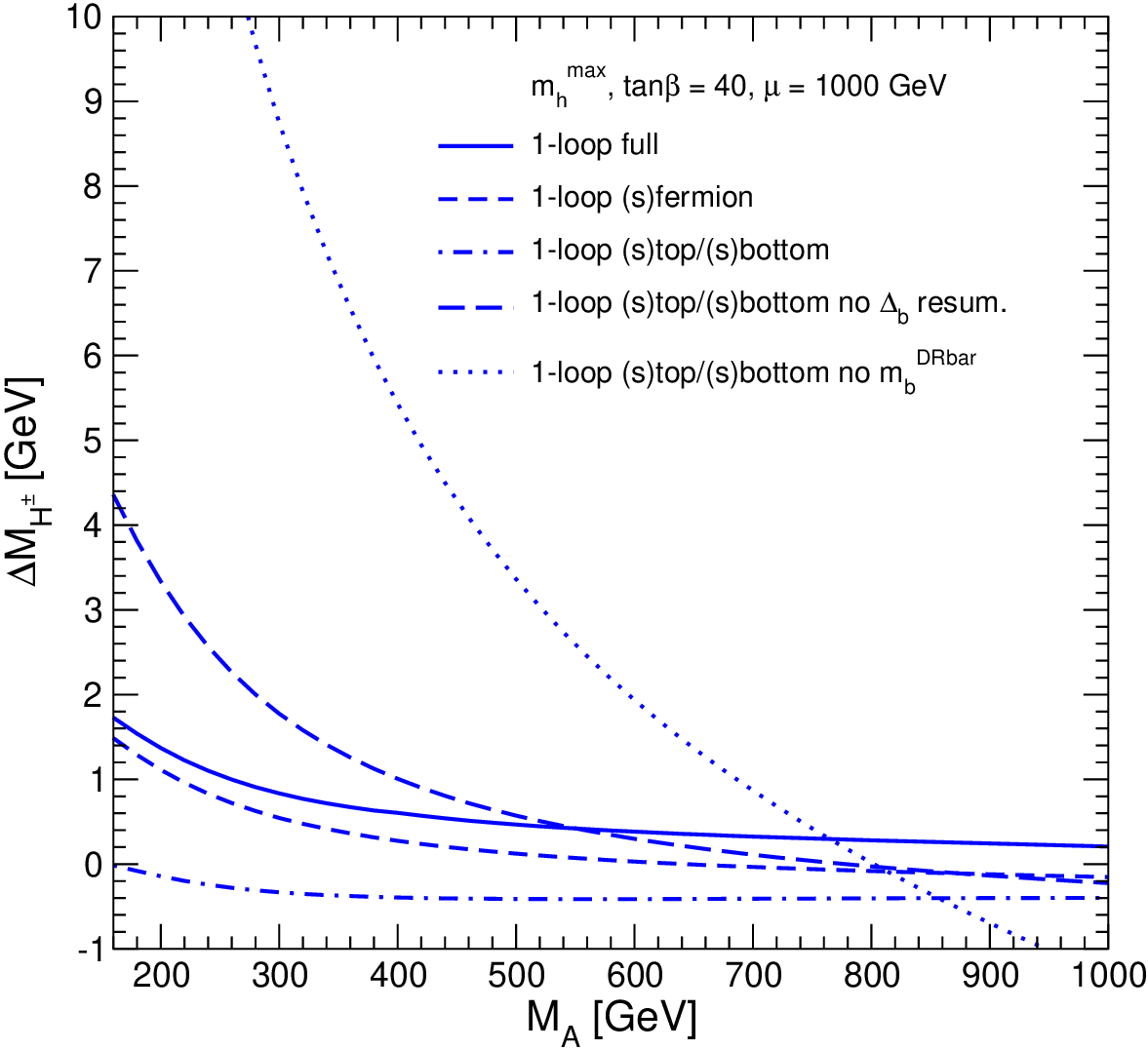}}
\fcaption{$\De\MHp := \MHp - \mHp$ is shown in the $\mhmax$ scenario as
  a function of $\MA$ for $\mu = 100 \gev$ (left) and $\mu = 1000 \gev$
  (right) and $\tb = 40$, evaluated at the one-loop level.
  We show the full one-loop result including $\db$ corrections
  (solid lines), the pure SM fermion/sfermion contribution (short
  dashed), the $\Stop/\Sbot$ contribution (dot-dashed), 
  the $\Stop/\Sbot$ corrections excluding the $\db$ corrections but
  using $\mbms$ (long dashed), and the $\Stop/\Sbot$ corrections
  excluding the $\db$ resummation and using the bottom pole mass,
  $\mb$ (dotted).
}
\label{fig:DeMHp_MA}
\end{figure}

\begin{figure}[htb!]
\vspace{2em}
\centerline{\includegraphics[width=0.5\linewidth]{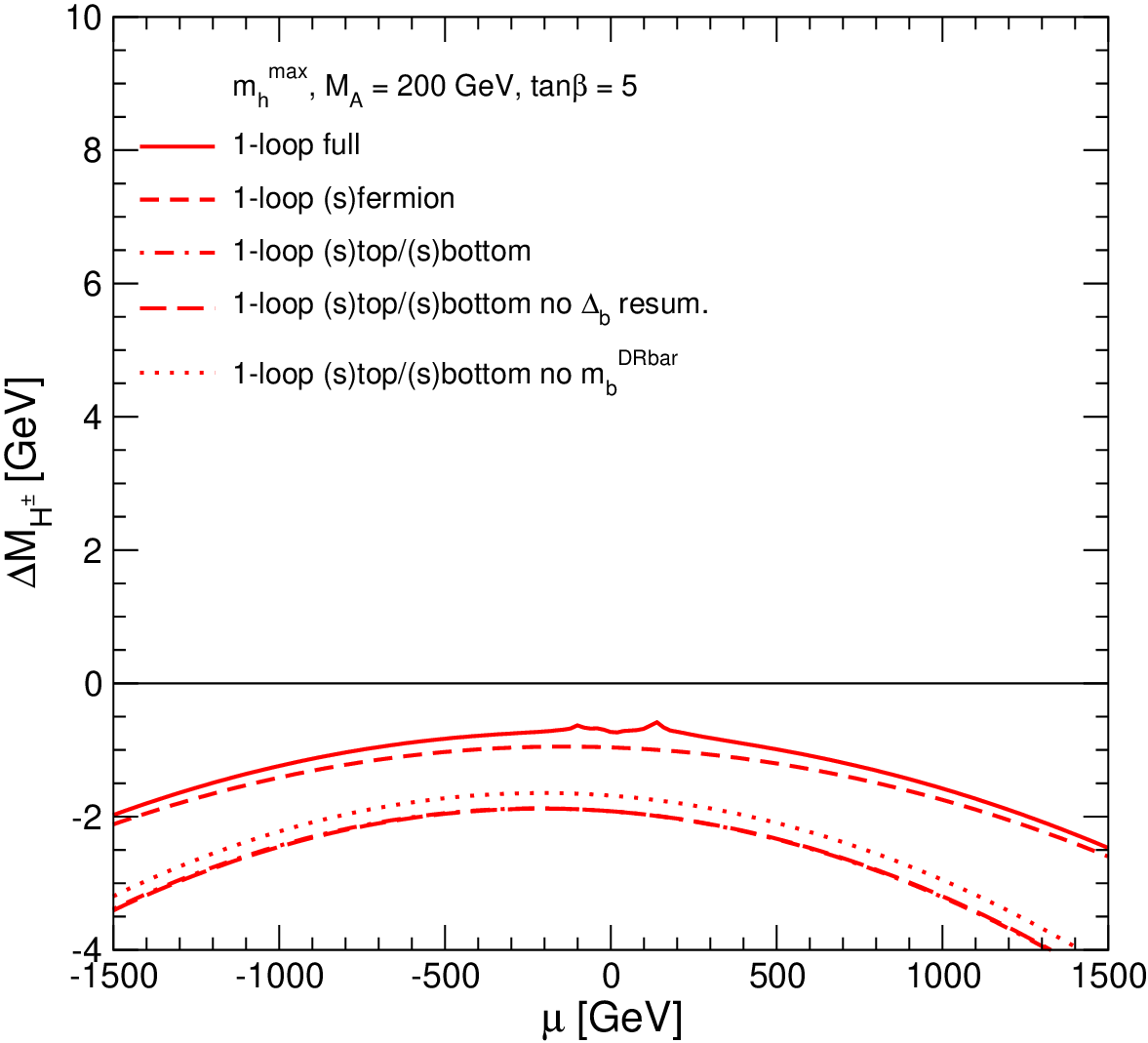}
            \hspace{1em}
            \includegraphics[width=0.5\linewidth]{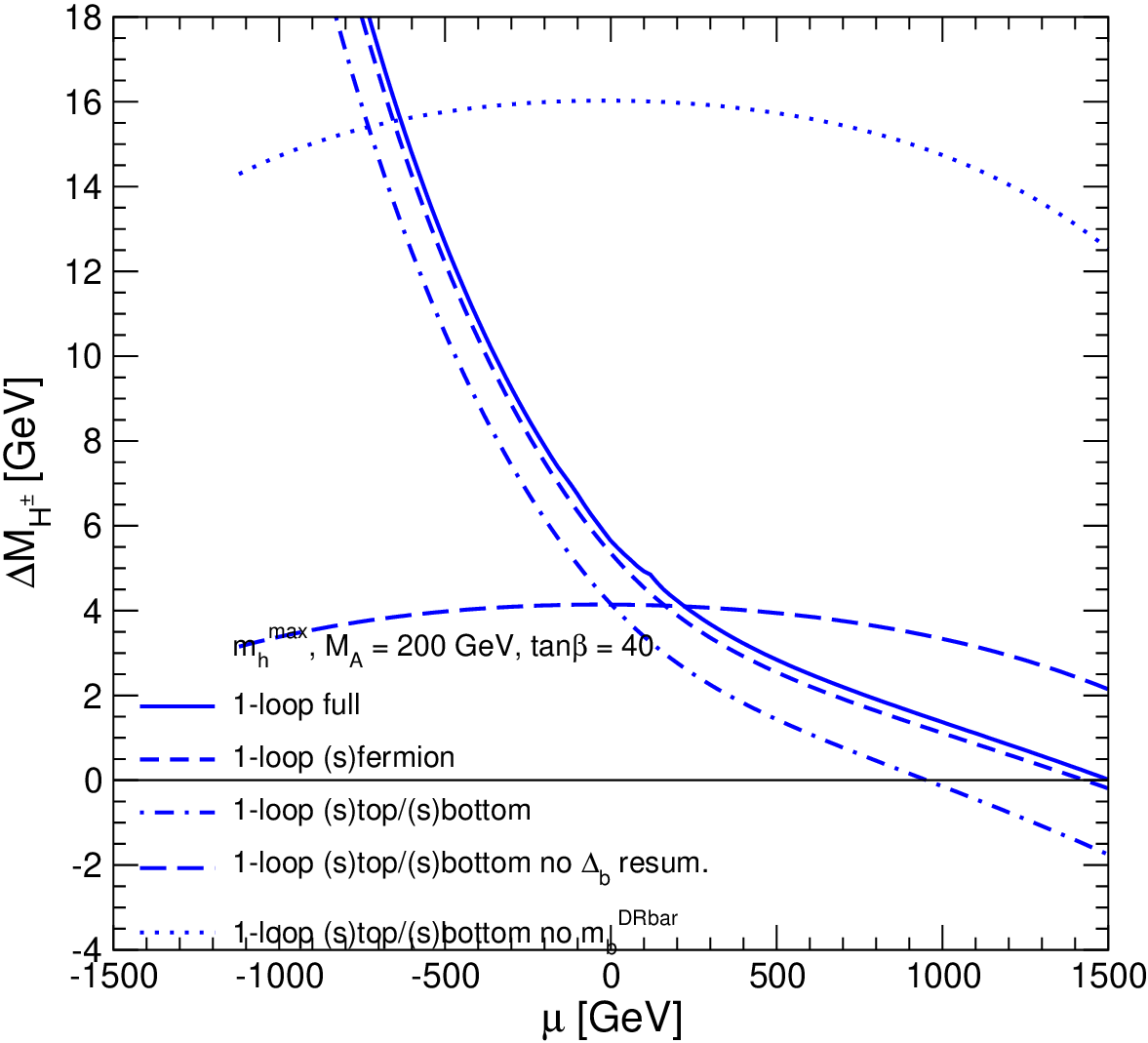}}
\fcaption{$\De\MHp := \MHp - \mHp$ is shown in the $\mhmax$ scenario as
  a function of $\mu$ for $\tb = 5$ (left) and $\tb = 40$ (right) 
  and $\MA = 200 \gev$, evaluated at the one-loop level.
  The line coding is as in \reffi{fig:DeMHp_MA}. 
}
\label{fig:DeMHp_mu}
\end{figure}

We start with the analysis of the various one-loop contributions. 
In \reffis{fig:DeMHp_MA} -- \ref{fig:DeMHp_mu}
we show $\De\MHp := \MHp - \mHp$, i.e.\ the
difference between the result with radiative corrections and the
tree-level value, in various approximations. 
The solid lines are the full one-loop result
including the $\db$ resummation, see \refeq{def:Deb}. The first
approximation to this is 
shown as short-dashed lines, where only the contributions from SM
fermions and their SUSY partners (i.e.\ all squarks and sleptons) are
taken into account, still including the $\db$ corrections. The next step
of approximation is shown as dot-dashed lines, where only corrections
from the $t/b$~and $\Stop/\Sbot$~sector are included, still with the
$\db$ resummation. The penultimate step of the approximation is to leave
out the $\db$ corrections, but using $\mbms$ (i.e.\ including the SM QCD 
corrections, see \refeq{mbdrbar}) in the
Higgs boson couplings, shown as the long-dashed lines. The final
step in  the approximation is to drop the SM QCD corrections, i.e.\
replacing $\mbms$ by the bottom pole mass, $\mb = 4.8 \gev$, 
in the Higgs Yukawa couplings, shown as the dotted lines.

First, in \reffis{fig:DeMHp_MA},
we analyze the dependence of $\MHp$ on $\MA$ in the $\mhmax$ scenario.
The left (right) plot of \reffi{fig:DeMHp_MA} shows $\De\MHp$ as a function
of $\MA$ for $\tb = 40$ and $\mu = 100 (1000) \gev$.
It should be noted that the very low $\MA$ values are by now ruled
out by the LHC heavy MSSM Higgs boson searches~\cite{CMSHiggsMSSM}
for this value of $\tb$.
However, in order to display the full parameter dependence
we do not include these bounds here.
The full result (solid lines) yields one-loop corrections between 
$1.5 \gev$ and $6.0 \gev$ for low $\MA$, becoming smaller for
increasing $\MA$. The still allowed $\MA$ values should give 
one-loop corrections of \order{2 \gev} in this scenario for small $\mu$.
The $f/\Sf$~sector (short-dashed) gives a rather good approximation, better
than $0.5 \gev$. Going to the $t/b/\Stop/\Sbot$ approximations
(dot-dashed) yields a prediction that differs from the full result
by up to $\sim 2 \gev$ for low
$\MA$. The $f/\Sf$ corrections besides the ones from third generation
squarks are roughly independent of the Yukawa couplings of the
various (s)fermions and are of pure electroweak type, and can grow as 
$\log(\msusy/\MW)$~\cite{mhp1lD}, and larger masses lead to larger
corrections. Consequently, taking into account only 
the third generation (s)quark contribution can yield non-negligible
uncertainties in the $\MHp$ prediction. In the next step the $\db$
corrections are neglected, which are formally beyond the one-loop order,
resulting in the long-dashed lines. The comparison between the
dot-dashed and the long-dashed lines shows that the impact of the
$\db$ corrections is small, below $\sim 500 \mev$ for $\mu = 100 \gev$,
but can be larger than $4 \gev$ for $\mu = 1000 \gev$, see
\refeq{def:Deb}. 
Finally we consider an approximation where the SM QCD corrections
to the bottom Yukawa coupling  
are dropped, i.e.\ $\mb$ is used instead of $\mbms$, resulting in the
dotted lines. These contributions can be larger than all other steps of
approximations in the region of large $\tb$ considered here.
Neglecting the SM QCD corrections in $\mb$ shifts $\De\MHp$
upwards by more than $10 \gev$, depending on the scenario.

\smallskip
In order to analyze the dependence of the $\MHp$ prediction on $\mu$ we
show in \reffi{fig:DeMHp_mu} $\De\MHp$ in the $\mhmax$ scenario 
as a function of $\mu$ for $\MA = 200 \gev$ and $\tb = 5 (40)$ in the
left (right) plot.
Again, the large $\tb$ values are by now experimentally excluded by
the LHC heavy MSSM Higgs searches for this value of
$\MA$~\cite{CMSHiggsMSSM}, but the two 
``extreme'' $\tb$ values are meant to give an idea about the
possible variations. 
We start with the case of $\tb = 5$, see the left plot
of \reffi{fig:DeMHp_mu}.
The $t/b/\Stop/\Sbot$ corrections neglecting the SM QCD corrections
(dotted line) are
nearly symmetric in $\mu$, ranging between $-2$ and $-4 \gev$. 
Including the SM QCD corrections (long-dashed) has a negligible
impact. The same holds for the $\db$ corrections (dashed-dotted) due
to the small value of $\tb$, and the two lines lie on top of each
other.  Including the full (s)fermion corrections, on the other hand, 
has a sizable impact on the result. The contributions 
from the other s/fermions partially
cancel the $t/b/\Stop/\Sbot$ corrections. Including also the
non-(s)fermionic contributions yields a total one-loop effect that stays
below $\sim -2 \gev$. 

The results look quite different for $\tb = 40$ as shown in the right
plot of \reffi{fig:DeMHp_mu}. 
For negative $\mu$, the enhancement of the bottom Yukawa coupling 
can become very strong due to the large $\tb$ value. 
In the $\mhmax$ scenario $\mu \lsim -1200 \gev$ yields 
$\db \to -1$, i.e.\ the model enters the non-perturbative regime, and no
evaluations in the Higgs sector are possible. Consequently, the
corresponding curves in the right plot of \reffi{fig:DeMHp_mu}
stop at $\mu \approx -1100 \gev$. 
The pure $t/b/\Stop/\Sbot$
corrections (dotted line) reach 13--$16 \gev$
if they are evaluated with the bottom pole mass.
Including the SM QCD corrections (long-dashed) into the effective
bottom quark mass strongly reduces the effect to the level of 
2--$4 \gev$. In the next step the $\db$ effects are included (dot-dashed
line). Due to $\db \propto \mu\,\tb$ the inclusion of $\db$ results in a
strong asymmetry of $\De\MHp$ with a larger correction to $\MHp$
for negative 
$\mu$ (corresponding to an enhanced bottom Yukawa coupling) and a much
smaller correction for positive $\mu$ (corresponding to a suppressed bottom
Yukawa coupling). 
Including the full one-loop corrections 
the overall correction in the $\mhmax$ scenario ranges from 
$\De\MHp \gsim 18 \gev$ for $\mu \lsim -1000 \gev$ to
$\De\MHp \approx 0$ for $\mu = +1500 \gev$.

\begin{figure}[htb!]
\vspace{1em}
\centerline{\includegraphics[width=0.5\linewidth]{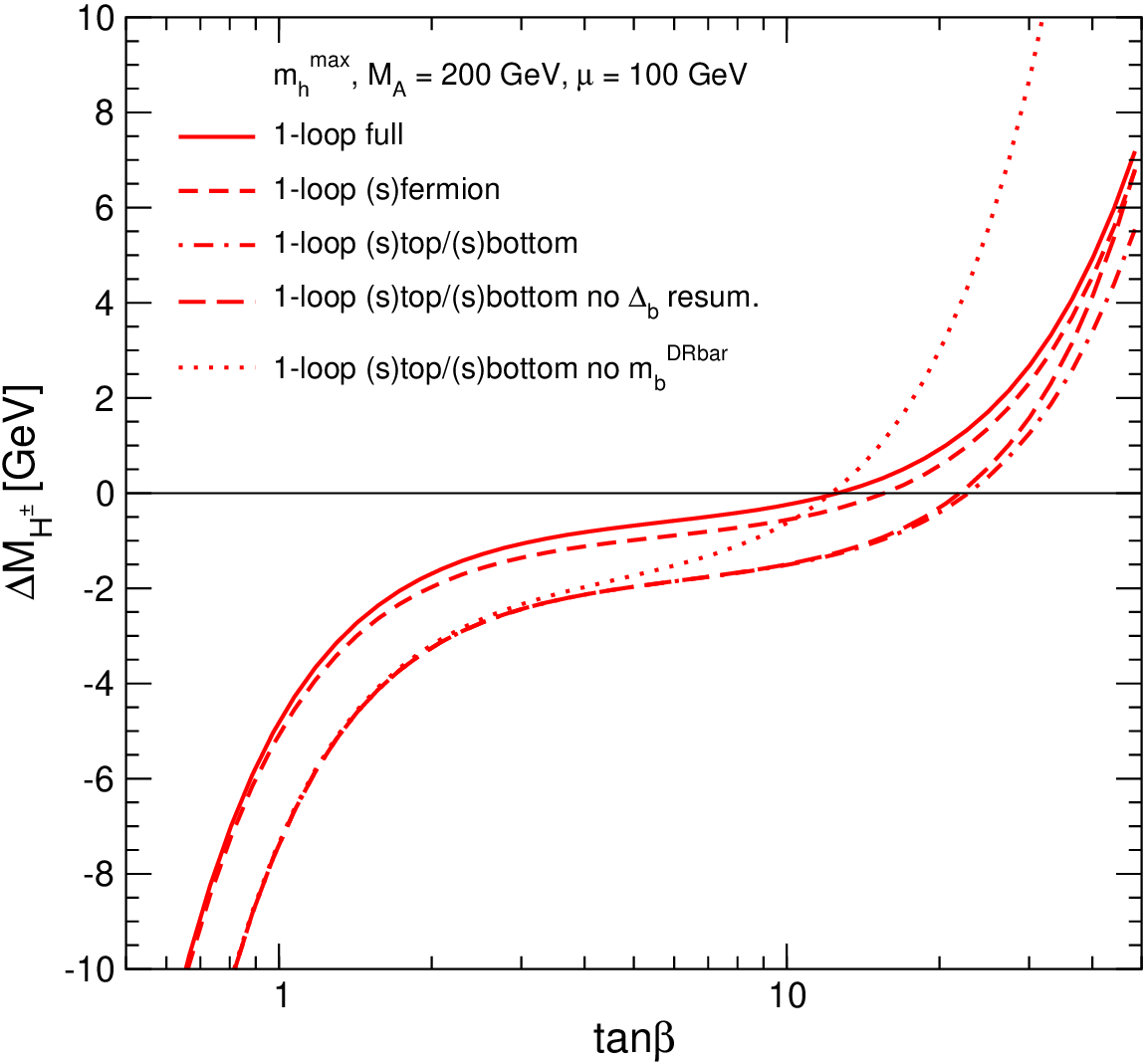}
            \hspace{1em}
            \includegraphics[width=0.5\linewidth]{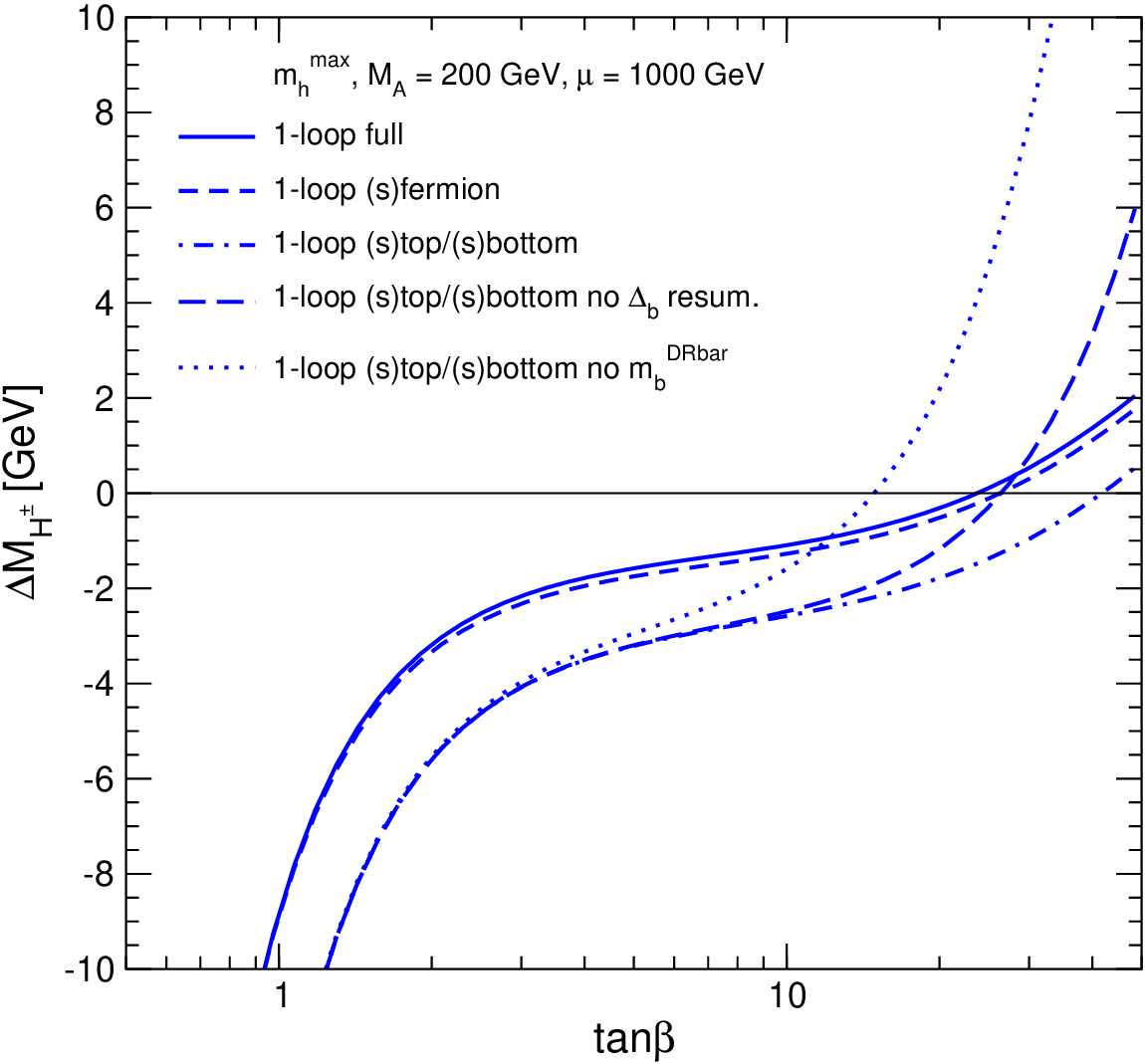}}
\fcaption{$\De\MHp := \MHp - \mHp$ is shown in the $\mhmax$ scenario
  as a function of $\tb$ for $\mu = 100 \gev$ (left) and $\mu = 1000 \gev$
  (right) and $\MA = 200 \gev$, evaluated at the one-loop level.
  The line coding is as in \reffi{fig:DeMHp_MA}. 
}
\label{fig:DeMHp_tb}
\vspace{1em}
\end{figure}

\smallskip
The dependence on $\tb$ is analyzed in \reffi{fig:DeMHp_tb}.
We show $\De\MHp$ in the $\mhmax$ scenario as a function of $\tb$
for $\MA = 200 \gev$ and as before for $\mu = 100 (1000) \gev$ in the
left (right) plot. 
It should be noted that values of $\tb$ around~1 are
excluded by LEP Higgs searches~\cite{LEPHiggsMSSM}, whereas large values
are excluded by LHC Higgs searches for this value of
$\MA$~\cite{CMSHiggsMSSM}. 
The sign and size of the one-loop correction to $\MHp$ depends strongly
on $\tb$, which enters the Higgs couplings to (s)fermions as well as 
the $\db$ corrections. Negative corrections occur for 
$\tb \lsim 10$, while
positive values of $\De\MHp$ are obtained for large $\tb$ values.
In the phenomenologically allowed region of $\tb$ the corrections
stay within $\De\MHp = \pm 2 \gev$.
As in the plots of \reffis{fig:DeMHp_MA}, the effect of the non-sfermion
sector in comparison with the $f/\Sf$ contributions (short-dashed
lines) is relatively small and stays below $0.5 \gev$.  
The Yukawa coupling independent effects (dot-dashed lines) are 
$\sim 2 \gev$, largely independent of $\tb$. The contribution from the $\db$
effects is negligible for $\tb \lsim 5$ and grows with increasing $\tb$,
reaching several GeV for large $\tb$ and $\mu = 1000 \gev$. 
On the other hand, for $\mu = 100 \gev$ these corrections stay very small 
even for the largest $\tb$ values.
The biggest effects again can arise from the inclusion of the SM QCD
corrections to $\mb$ for $\tb \gsim 5$. Largely independently of the
scenario and the choice for $\mu$ they reach 5--$10 \gev$.

\begin{figure}[htb!]
\vspace{1em}
\centerline{\includegraphics[width=0.5\linewidth]{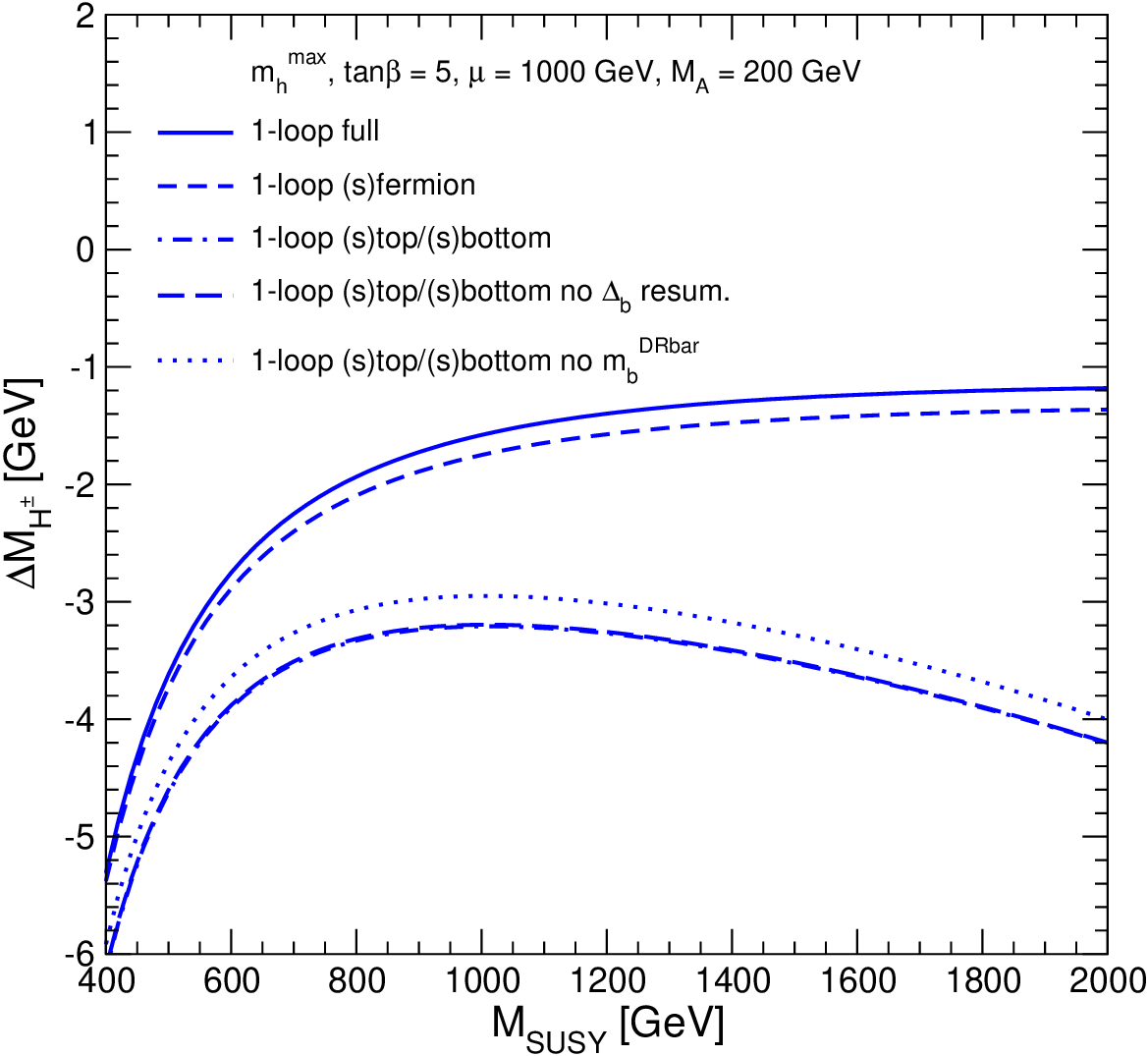}
            \hspace{1em}
            \includegraphics[width=0.5\linewidth]{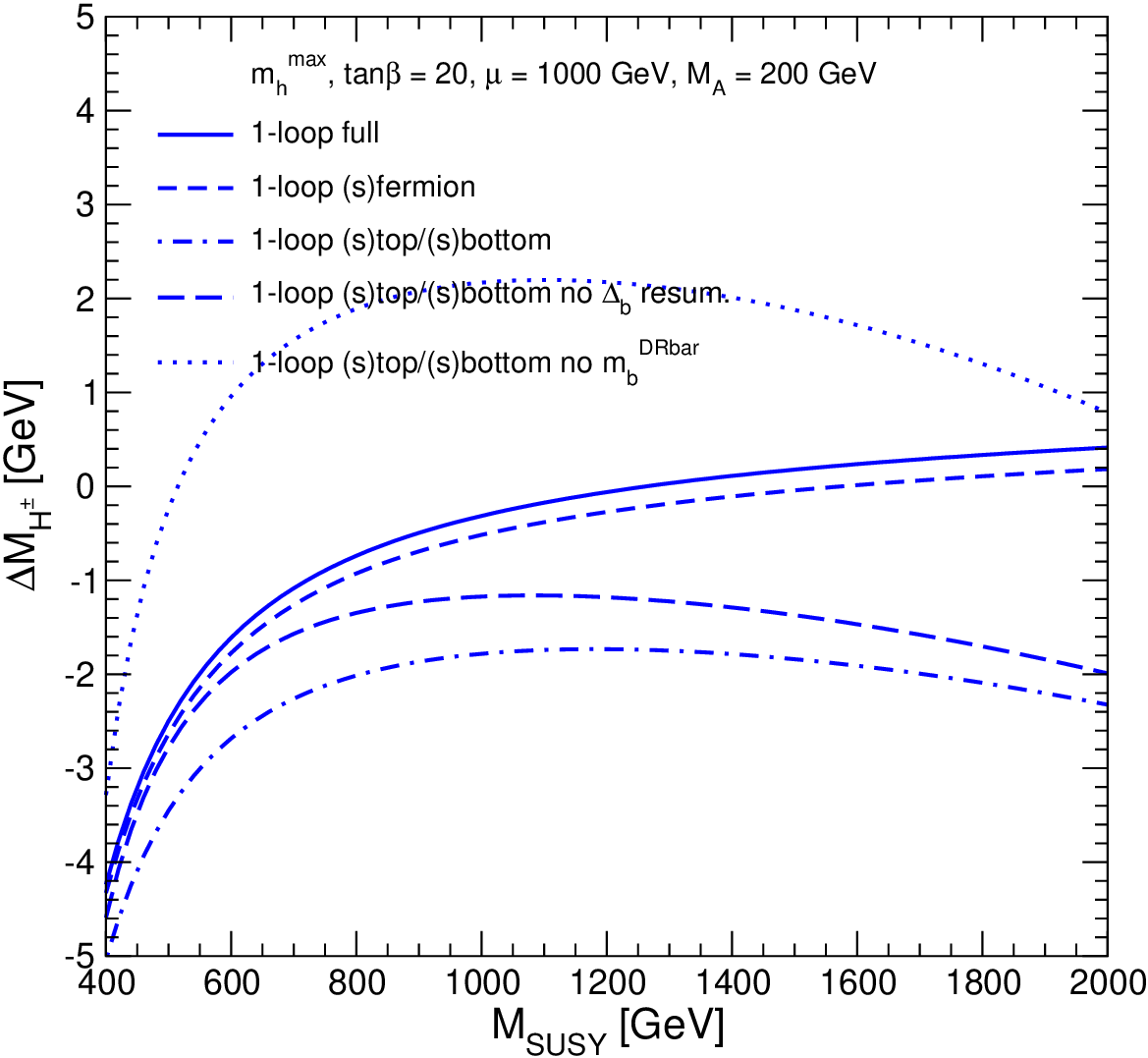}}
\fcaption{$\De\MHp := \MHp - \mHp$ is shown in the $\mhmax$ scenario
  as a function of $\msusy$ for $\tb = 5$ (left) and $\tb = 20$
  (right), $\MA = 200 \gev$ and $\mu = 1000 \gev$, evaluated at
  the one-loop level. 
  The line coding is as in \reffi{fig:DeMHp_MA}. 
}
\label{fig:DeMHp_Msusy}
\vspace{1em}
\end{figure}

\smallskip
Next, in \reffi{fig:DeMHp_Msusy} we show the dependence on $\msusy$.
The SUSY mass scale (which we chose to be equal for all 
sfermions, see above)
enters via contributions $\propto \log(\msusy/\MW)$ or
$\propto \MW^2/\msusy^2$ into the charged
Higgs-boson mass prediction~\cite{mhp1lD}, where several competing
contributions have been identified. One is proportional to
large Yukawa couplings from the top/bottom sector, while another
one stems from the electroweak couplings of scalar fermions and is
similar for all flavors.

In the left plot of \reffi{fig:DeMHp_Msusy} we show $\De\MHp$ as a
function of $\msusy$ in the $\mhmax$ scenario for $\mu = 1000 \gev$,
$\MA = 200 \gev$ and $\tb = 5$. One can see that the $b/\Sbot$
contributions, which are influenced strongly by the bottom
Yukawa coupling and the $\db$~corrections, do not play a prominent role
as they change $\De\MHp$ only weakly for small $\tb$. The
contributions from the lighter 
fermions (i.e.\ neither top nor bottom) and their SUSY partners, on the
other hand, become very important for $\msusy \gsim 1000 \gev$. Without
those corrections (dot-dashed line) rather large negative
  contributions to $\MHp$ would occur for large $\msusy$, 
while including these corrections (short dashed line) 
$\De\MHp$ flattens out for large $\msusy$, 
reaching $\sim -1 \gev$ at $\msusy = 2000 \gev$. The
corrections from the non-(s)fermion sector are small and change $\MHp$
by less than about $0.2 \gev$. 
A qualitatively similar behavior can be observed for $\tb = 20$ 
(which is close to the current sensitivity limits of heavy MSSM
Higgs searches at the LHC~\cite{CMSHiggsMSSM}) as shown
in the right plot of \reffi{fig:DeMHp_Msusy}. Due to the larger value of
$\tb$ the $b/\Sbot$~corrections and $\db$~effects are much more
pronounced. The contributions from the (s)fermion sector beyond
$t/\Stop/b/\Sbot$ are sizable for $\msusy \gsim 1000 \gev$. Due to
numerical cancellations the full one-loop correction to $\MHp$ is close
to zero for this part of the SUSY parameter space. This is in agreement
with the right plot of \reffi{fig:DeMHp_tb}. 
In summary, for large $\msusy$ especially the corrections from the 
{\em full} (s)fermion sector have to be taken into account.

\begin{figure}[htb!]
\vspace{1em}
\centerline{\includegraphics[width=0.85\linewidth]{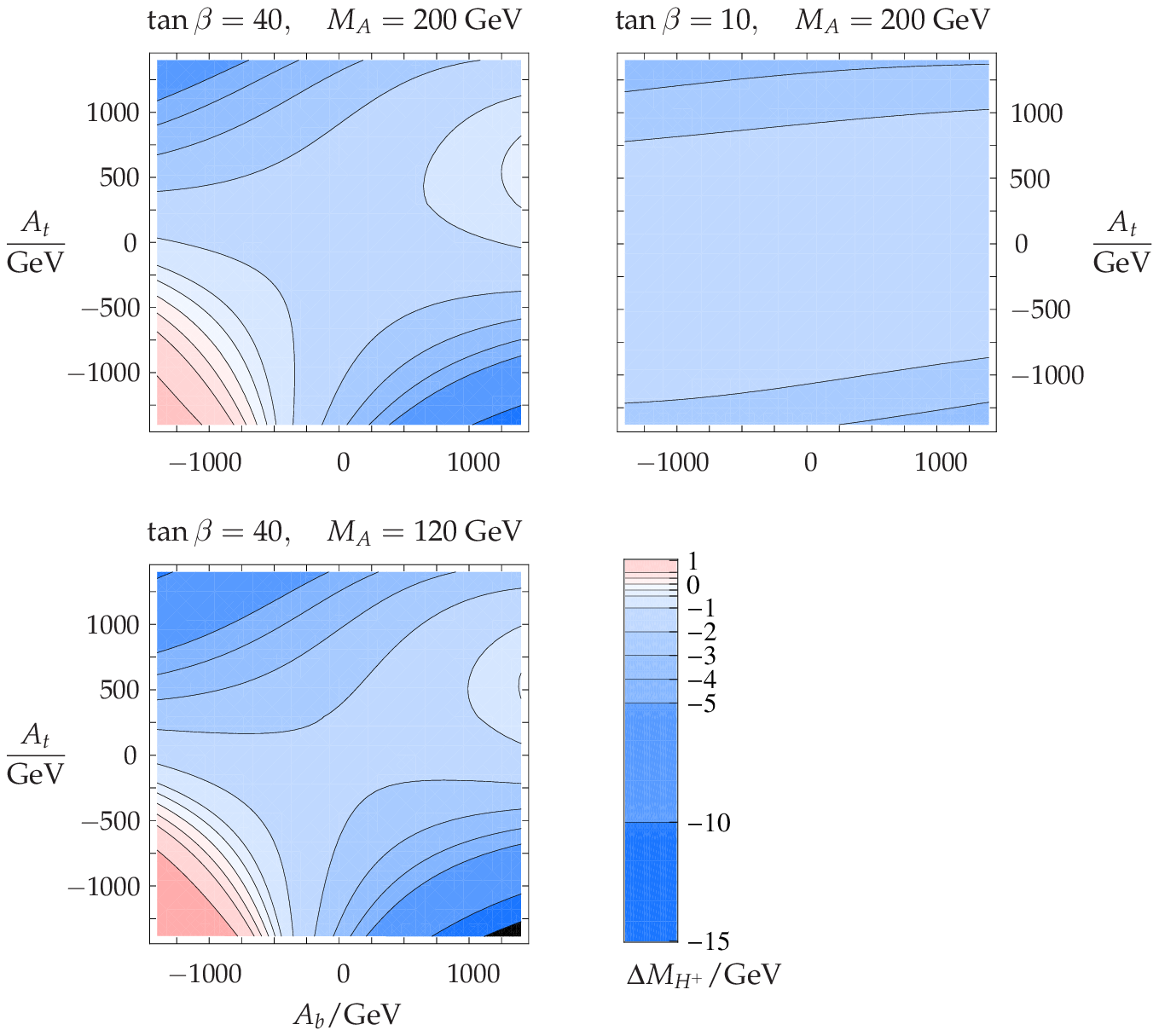}}
\fcaption{The size of the full one-loop correction
$\De\MHp := \MHp - \mHp$ is shown in the $\Ab$--$\At$ plane
 for $\MA = 200 (120) \gev$ in the top (bottom) row and $\tb = 40 (10)$
 in the left (right) column. The other parameters are $\msusy = 500 \gev$, 
 $\mu = 1000 \gev$, $M_2 = 500 \gev$.
}
\label{fig:DeMHp_AtAb}
\vspace{1em}
\end{figure}

\smallskip
We finally analyze the size of the full one-loop corrections in the
case of 
$\At \neq \Ab$ in \reffi{fig:DeMHp_AtAb}. We show the results in the 
$\Ab$--$\At$ plane for $\MA = 200 (120) \gev$ in the top (bottom) row
and $\tb = 40 (10)$ in the left (right) column. 
Again, the ``extreme'' choices for $\MA$ and $\tb$, partially
excluded by LHC Higgs searches~\cite{CMSHiggsMSSM} indicate the range
of the possible size of the corrections.
The other parameters are
$\msusy = 500 \gev$, $\mu = 1000 \gev$, $M_2 = 500 \gev$. 

The color coding in the plots 
tells the value of $\De\MHp$. At small $\tb$, as can be seen in
the upper right plot, the value of $\De\MHp$ depends mainly on $\At$,
i.e.\ the convention $\Ab = \At$ in the $\mhmax$ scenario
does not have a relevant impact on the $\MHp$ evaluation at the one-loop
level for small $\tb$. 
This changes for large $\tb$ as can be observed in the two left
plots of the figure. The main diagonal corresponds to $\At = \Ab$ and
exhibits (for both $\MA$ values) relatively small corrections up to 
$\sim 3 \gev$. The other extreme, $\At = -\Ab$, on the other hand,
yields much larger corrections, exceeding $\De\MHp = 10 \gev$
for large $|\At|$.
Consequently, a full one-loop calculation, allowing for different values
of $\At$ and $\Ab$ is crucial for a precise $\MHp$ evaluation.


\subsection{Two-loop corrections}
\label{sec:twoloop}

We now turn to the analysis of the effects of the two-loop corrections of
\order{\alt\als}, where in the plots we denote ``2-loop'' as the full
one-loop corrections supplemented by the \order{\alt\als} contributions. 
As described in \refse{sec:mhp} we derived the
\order{\als} corrections to the one-loop \order{\mt^4/\MW^2} term.
In our numerical analysis we concentrate on cases where on the one hand
the full one-loop contribution to $\MHp$ is sizable, and on the other
hand the \order{\mt^4/\MW^2} corrections yield a relatively good
approximation to the full one-loop result. For these cases it can be
expected that the \order{\alt\als} two-loop corrections also constitute
a substantial part of the full two-loop contributions.

We focus here on relatively low $\tb$, since it is known that at large
$\tb$ the bottom/sbottom one-loop corrections are sizable (see the
previous subsection) and the \order{\alt\als} terms cannot be expected
to capture a leading piece of the two-loop contributions.
As can be seen in \refeqs{nomixfinal}, (\ref{maxmixfinal}), the
\order{\mt^4/\MW^2} terms going $\sim \mu$ are enhanced with
$\tb$. Therefore we present the two-loop \order{\alt\als} corrections as
a function of $\mu$. 
We furthermore set $\MA = 200 \gev$, which allows relatively large
absolute higher-order corrections.
The chosen parameters are thus mostly in agreement with the LHC
heavy MSSM Higgs searches (and we will not address this issue
in the rest of this subsection).

\begin{figure}[htb!]
\vspace{1em}
\centerline{\includegraphics[width=0.5\linewidth]{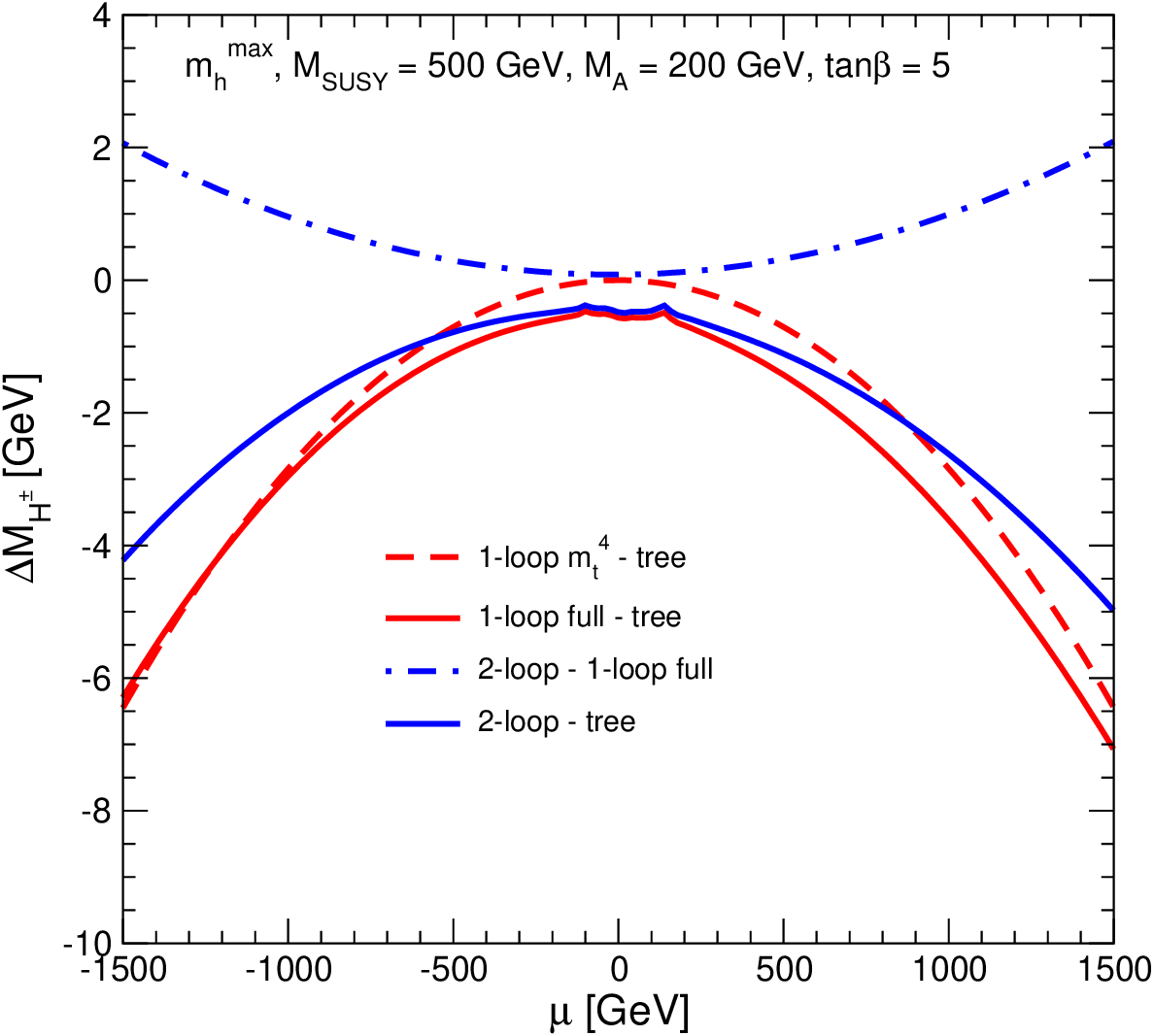}
            \hspace{1em}
            \includegraphics[width=0.5\linewidth]{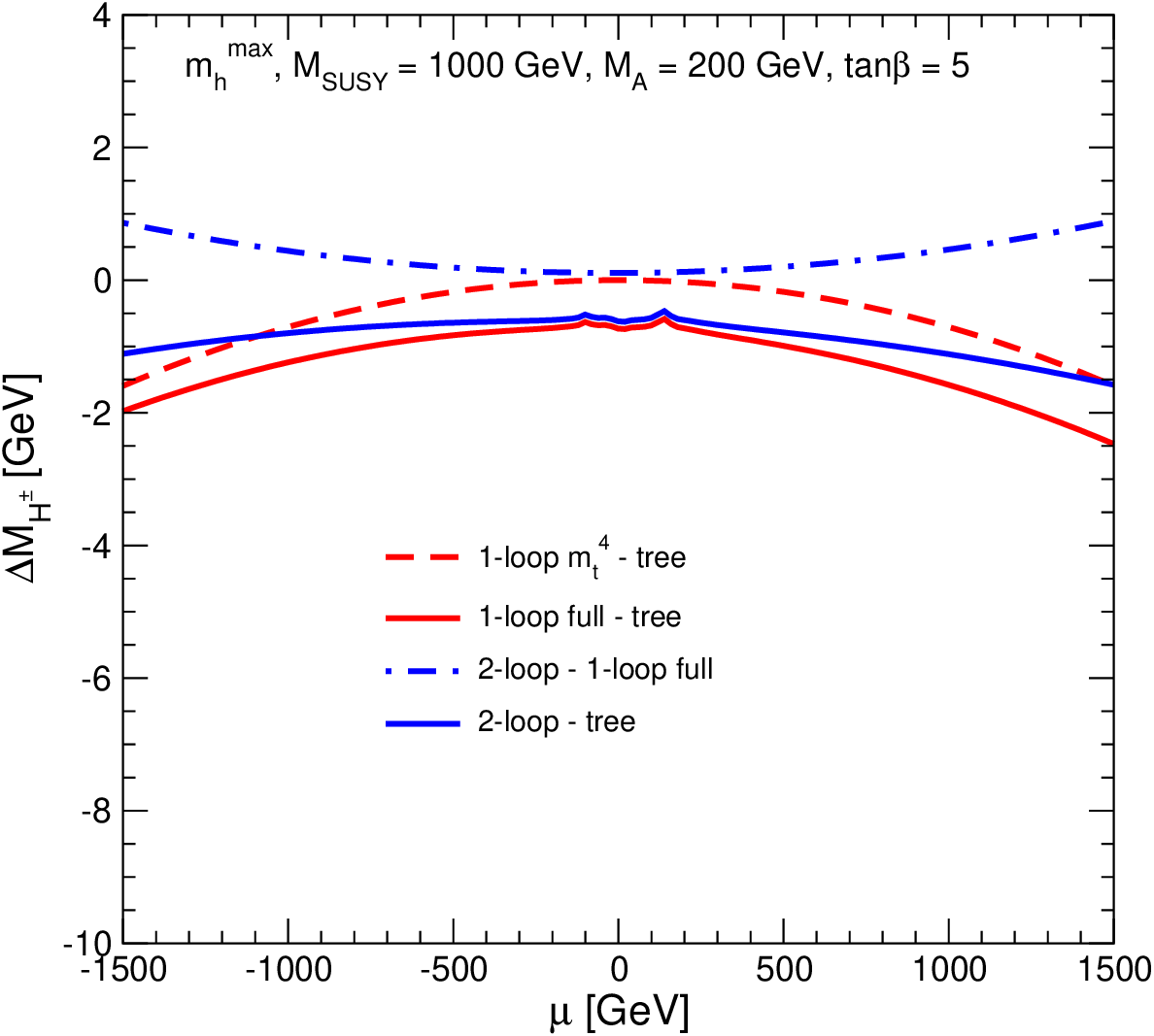}}
\fcaption{$\De\MHp := \MHp - \mHp$ 
  and $\MHp^{2-{\rm loop}} - \MHp^{1-{\rm loop}}$ are
  shown for $\MA = 200 \gev$ and  $\tb = 5$ as
  a function of $\mu$ in the $\mhmax$ scenario. 
  $\msusy$ is set to $500 \gev$ (left) and to $1000 \gev$ (right plot). 
  $\MHp$ is evaluated at
  the two-loop level (blue/dark gray solid).
  Also shown are the corresponding one-loop results of
  \order{\mt^4/\MW^2} (dashed), the full one-loop
  corrections (red/light gray solid) and the difference between the two-loop
  and the full one-loop result (dot-dashed).
}
\label{fig:DeMHp2L_tb5}
\vspace{1em}
\end{figure}

In \reffi{fig:DeMHp2L_tb5} we present $\De\MHp := \MHp - \mHp$ in the
$\mhmax$ scenario for $\MA = 200 \gev$ and $\tb = 5$ 
as a function of $\mu$ for $\msusy = 500 (1000) \gev$ in the left
(right) plot. $\MHp$ is evaluated including the
\order{\alt\als} corrections and shown as the blue/dark gray solid line. 
Also shown are the corresponding one-loop results of
\order{\mt^4/\MW^2} (dashed line), the full one-loop
corrections (red/light gray solid line) and the difference between the
two-loop and the full one-loop result (dot-dashed line).
Starting with the left plot, where we have set $\msusy = 500 \gev$,
we find that the full one-loop corrections are well
approximated by the \order{\mt^4/\MW^2} term. As expected for $\tb = 5$,
the $\db$ corrections do not play a prominent role and $\De\MHp$ appears
nearly symmetric for positive and negative $\mu$. 
The corresponding two-loop corrections modify the full one-loop result
by up to $\sim 2 \gev$ for $|\mu| \sim 1500 \gev$, i.e.\ the 
\order{\alt \als} corrections can be sizable in this case. 
A similar behavior can be observed in the right plot of
\reffi{fig:DeMHp2L_tb5}, where we have set $\msusy = 1000 \gev$
(i.e.\ as in the original definition of the $\mhmax$ scenario,
\refeq{mhmax}.) As expected, the absolute corrections to $\MHp$ turn out
to be smaller, see also \reffi{fig:DeMHp_Msusy}, and the two-loop terms
contribute up to $\sim 1 \gev$ for $|\mu| \sim 1500 \gev$ (where our
plot stops).

\begin{figure}[htb!]
\vspace{2em}
\centerline{\includegraphics[width=0.5\linewidth]{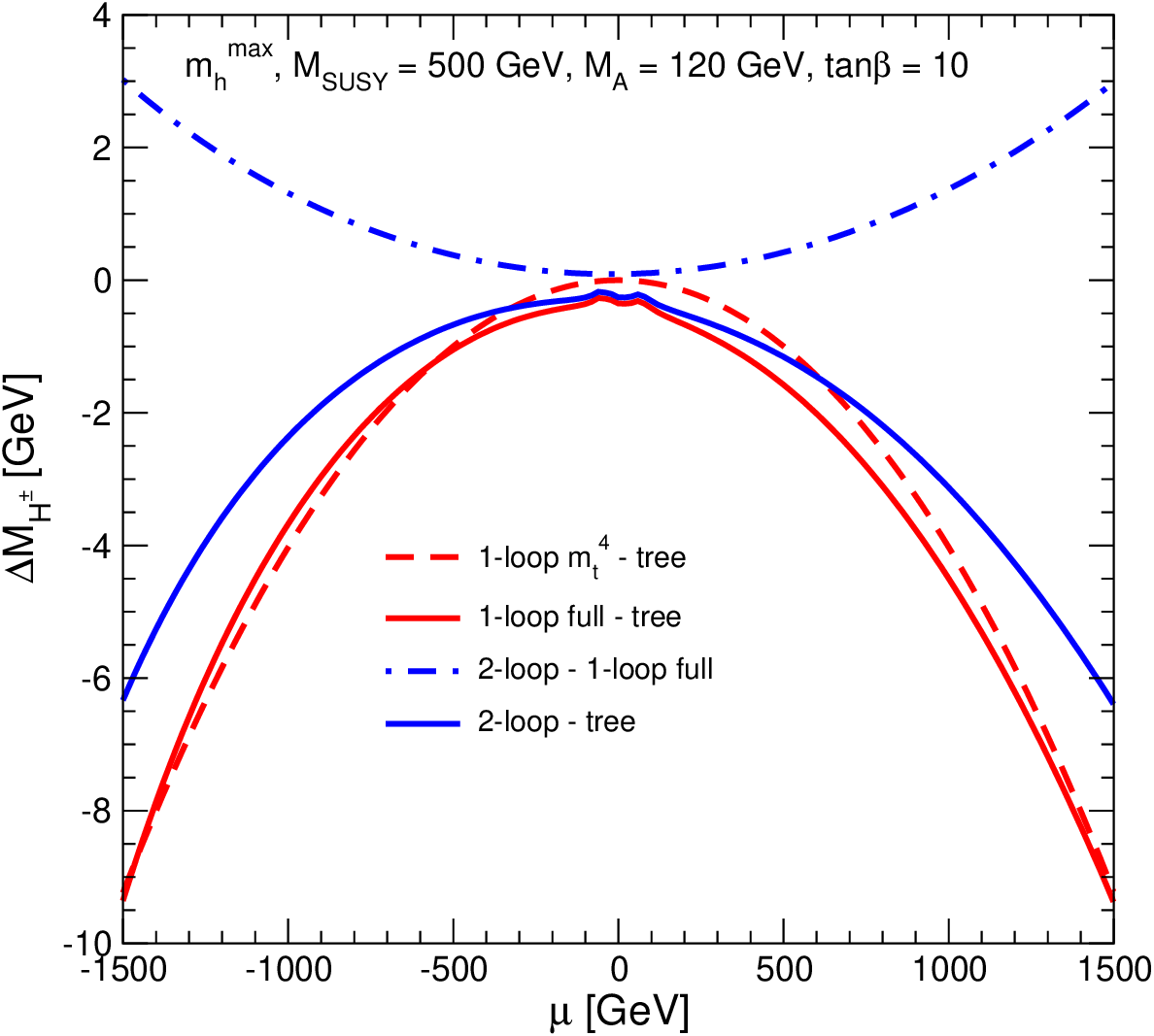}
            \hspace{1em}
            \includegraphics[width=0.5\linewidth]{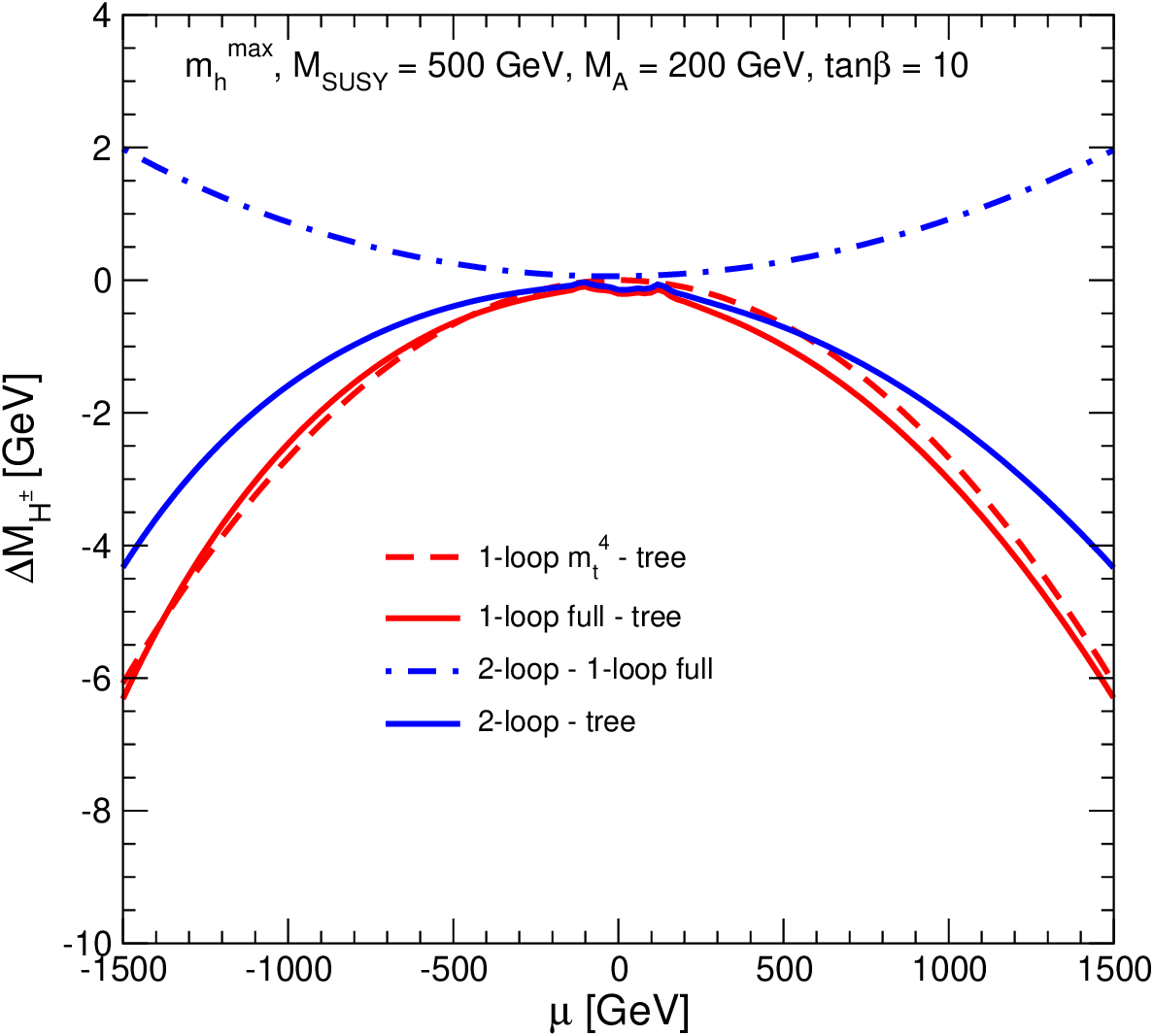}}
\fcaption{$\De\MHp := \MHp - \mHp$ 
  and $\MHp^{2-{\rm loop}} - \MHp^{1-{\rm loop}}$ are
  shown for $\MA = 120 \gev$ (left)
  and $\MA = 200 \gev$ (right plot), 
  $\tb = 10$ and $\msusy = 500 \gev$ as 
  a function of $\mu$ in the $\mhmax$ scenario.
  $\MHp$ is evaluated at
  the two-loop level (blue/dark gray solid).
  Also shown are the corresponding one-loop results of
  \order{\mt^4/\MW^2} (dashed), the full one-loop
  corrections (red/light gray solid) and the difference between the two-loop
  and the full one-loop result (dot-dashed).
}
\label{fig:DeMHp2L_tb10}
\end{figure}

For the remaining analysis we stick to the lower value of 
$\msusy = 500 \gev$, but go to somewhat larger $\tb$ values and
investigate also lower values of $\MA$. 
In \reffi{fig:DeMHp2L_tb10} we show $\De\MHp$ for $\tb = 10$ and 
$\MA = 120 (200) \gev$ in the left (right) plot. 
The results look qualitatively similar to the case of $\tb = 5$:
the $\mt^4/\MW^2$ approximation works well for the full one-loop
result. The two-loop corrections go up to $\sim 3 (2) \gev$ for large values
of $|\mu|$ for $\MA = 120 (200) \gev$.

\begin{figure}[htb!]
\vspace{2em}
\centerline{\includegraphics[width=0.5\linewidth]{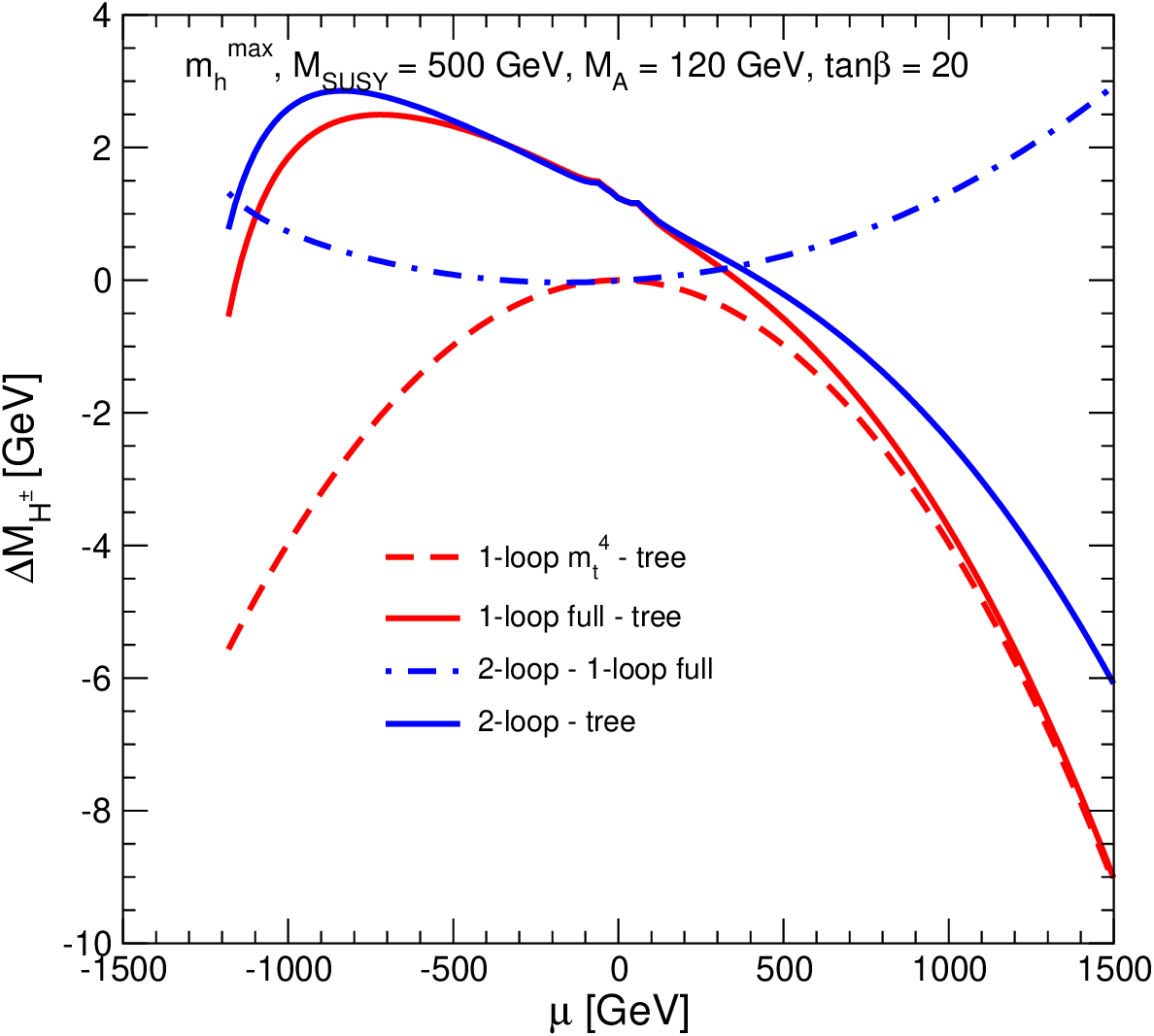}
            \hspace{1em}
            \includegraphics[width=0.5\linewidth]{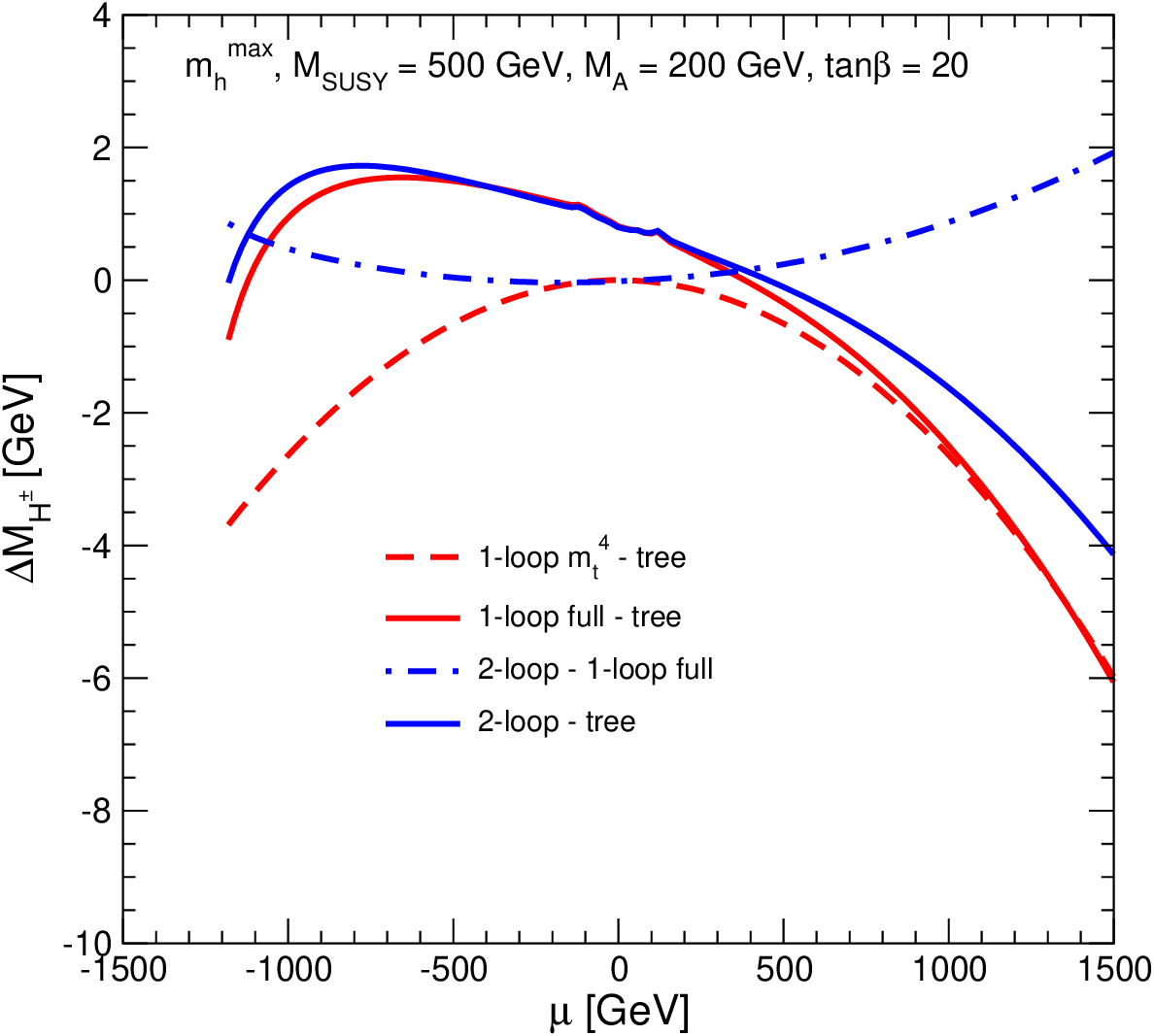}}
\fcaption{$\De\MHp := \MHp - \mHp$ 
  and $\MHp^{2-{\rm loop}} - \MHp^{1-{\rm loop}}$ are
  shown for $\MA = 120 \gev$ (left)
  and $\MA = 200 \gev$ (right plot), 
  $\tb = 20$ and $\msusy = 500 \gev$ as 
  a function of $\mu$ in the $\mhmax$ scenario.
  $\MHp$ is evaluated at
  the two-loop level (blue/dark gray solid).
  Also shown are the corresponding one-loop results of
  \order{\mt^4/\MW^2} (dashed), the full one-loop
  corrections (red/light gray solid) and the difference between the two-loop
  and the full one-loop result (dot-dashed).
}
\label{fig:DeMHp2L_tb20}
\vspace{-0.5em}
\end{figure}

In \reffi{fig:DeMHp2L_tb20} we go to even higher $\tb$ values and set
$\tb = 20$, where $\db$ effects are expected to become relevant. 
As for the previous figure we fix $\MA = 120 (200) \gev$ in the
left (right) plot. 
Sizable $\db$ effects can indeed be observed: for large
negative values of $\mu$, $\mu \lsim -1200 \gev$ the $\db$ corrections
become so large that an evaluation of the loop corrections to $\MHp$ was
not possible anymore (as was observed already in \reffi{fig:DeMHp_mu}). 
For negative $\mu$ the \order{\mt^4/\MW^2} corrections also 
start to deviate substantially from
the full one-loop result, and the corresponding
two-loop corrections cannot be expected to yield a good approximation to
the full two-loop result in the region of relatively large negative
$\mu$. 
For large and positive $\mu$, however, the $\mt^4/\MW^2$ approximation
works very well both for $\MA = 120 \gev$ (left plot) and
$\MA = 200 \gev$ (right plot), so that in this region the 
\order{\alt\als} corrections can be expected to provide a reasonable
approximation of the full two-loop corrections.
For $\mu = +1500 \gev$ the two-loop
corrections again are sizable and amount up to $\sim 2-3 \gev$.

\begin{figure}[htb!]
\centerline{\includegraphics[width=0.5\linewidth]{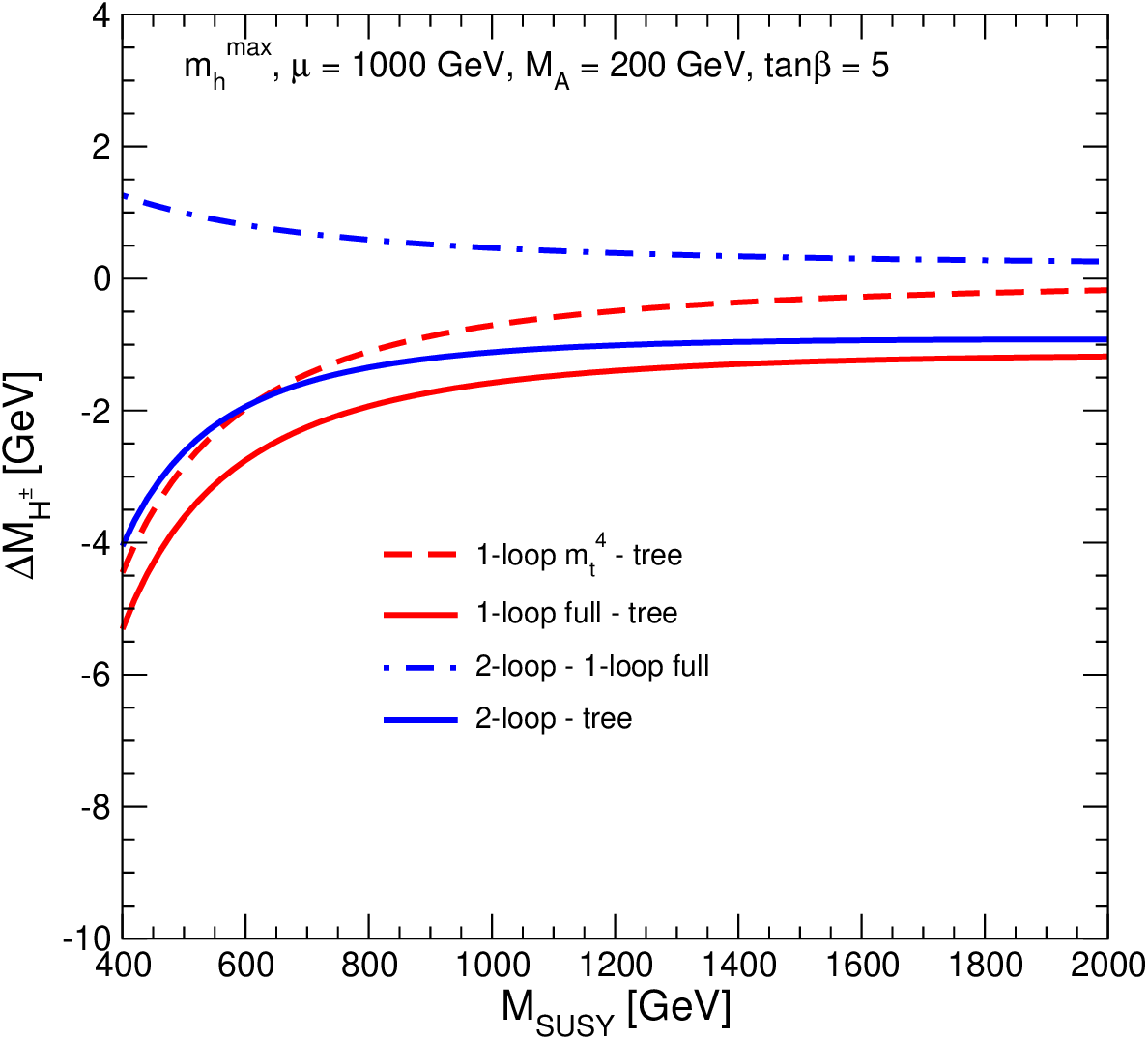}
            \hspace{1em}
            \includegraphics[width=0.5\linewidth]{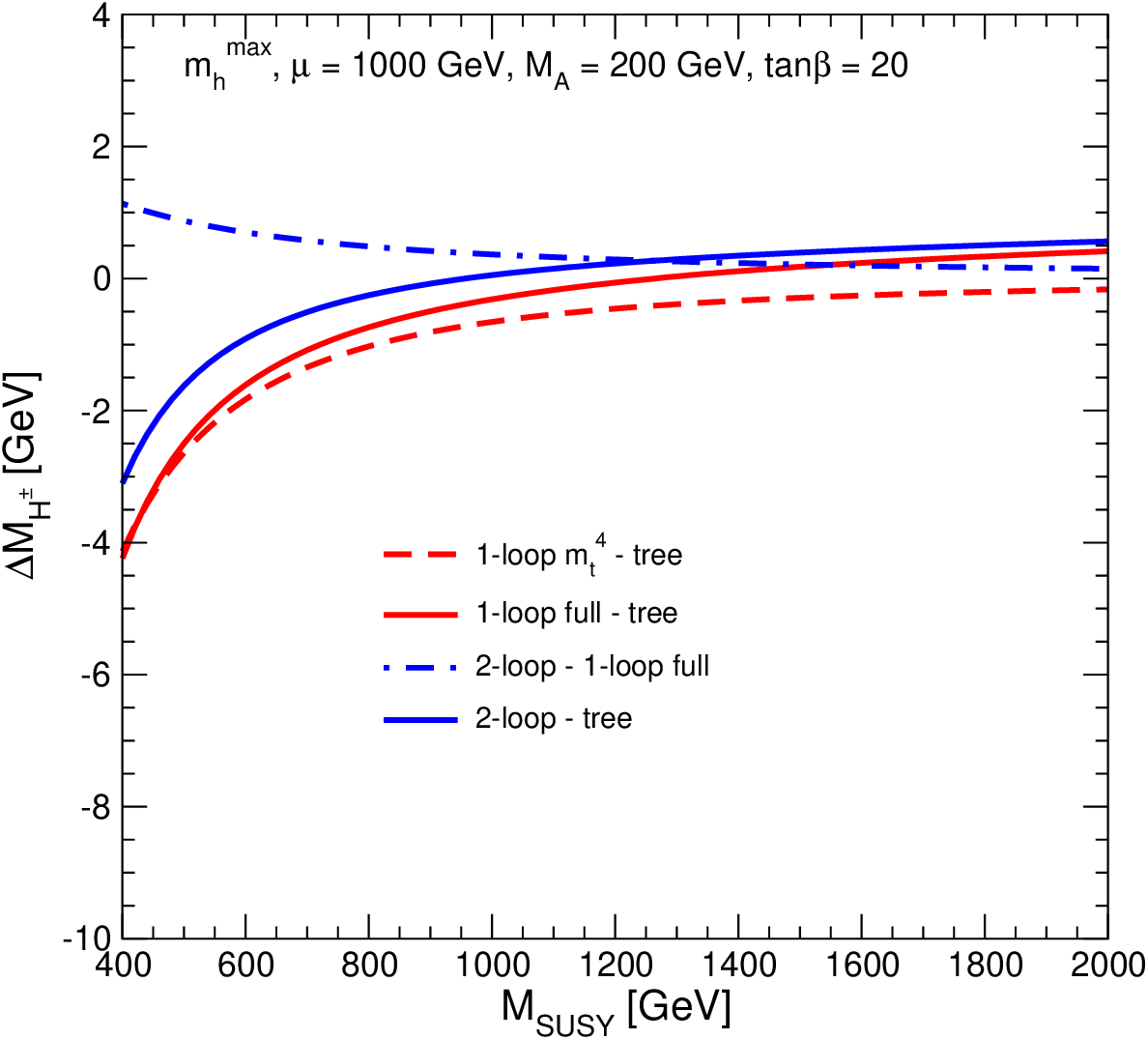}}
\fcaption{$\De\MHp := \MHp - \mHp$ 
  and $\MHp^{2-{\rm loop}} - \MHp^{1-{\rm loop}}$ are
  shown for $\MA = 200 \gev$, 
  $\tb = 5$ (left) and $\tb = 20$ (right plot) and $\mu = 1000 \gev$ as 
  a function of $\msusy$ in the $\mhmax$ scenario.
  $\MHp$ is evaluated at
  the two-loop level (blue/dark gray solid).
  Also shown are the corresponding one-loop results of
  \order{\mt^4/\MW^2} (dashed), the full one-loop
  corrections (red/light gray solid) and the difference between the two-loop
  and the full one-loop result (dot-dashed).
}
\label{fig:DeMHp2L_msusy}
\vspace{2em}
\end{figure}

We complete our two-loop analysis in the $\mhmax$ scenario with
\reffi{fig:DeMHp2L_msusy}, where 
we show $\De\MHp$ as a function of $\msusy$, in analogy to
\reffi{fig:DeMHp_Msusy}. In the left (right) plot we show the results
for $\tb = 5 (20)$ in the $\mhmax$ scenario (i.e.\ $\Xt = 2 \msusy$ and
$\mgl = 0.8 \msusy$) for $\mu = 1000 \gev$ and $\MA = 200 \gev$. 
The $\mt^4/\MW^2$ corrections approximate the full one-loop results
fairly well. The largest deviations occur for large values of $\msusy$,
where the other (s)fermion sectors become more
relevant, see the discussion on \reffi{fig:DeMHp_Msusy}.
For $\msusy = 400 \gev$, the
lowest value in our analysis, the two-loop \order{\alt\als} corrections
amount up to $\sim 1 \gev$. For large $\msusy$ this correction goes down
nearly to zero.  

\begin{figure}[htb!]
\centerline{\includegraphics[width=0.5\linewidth]{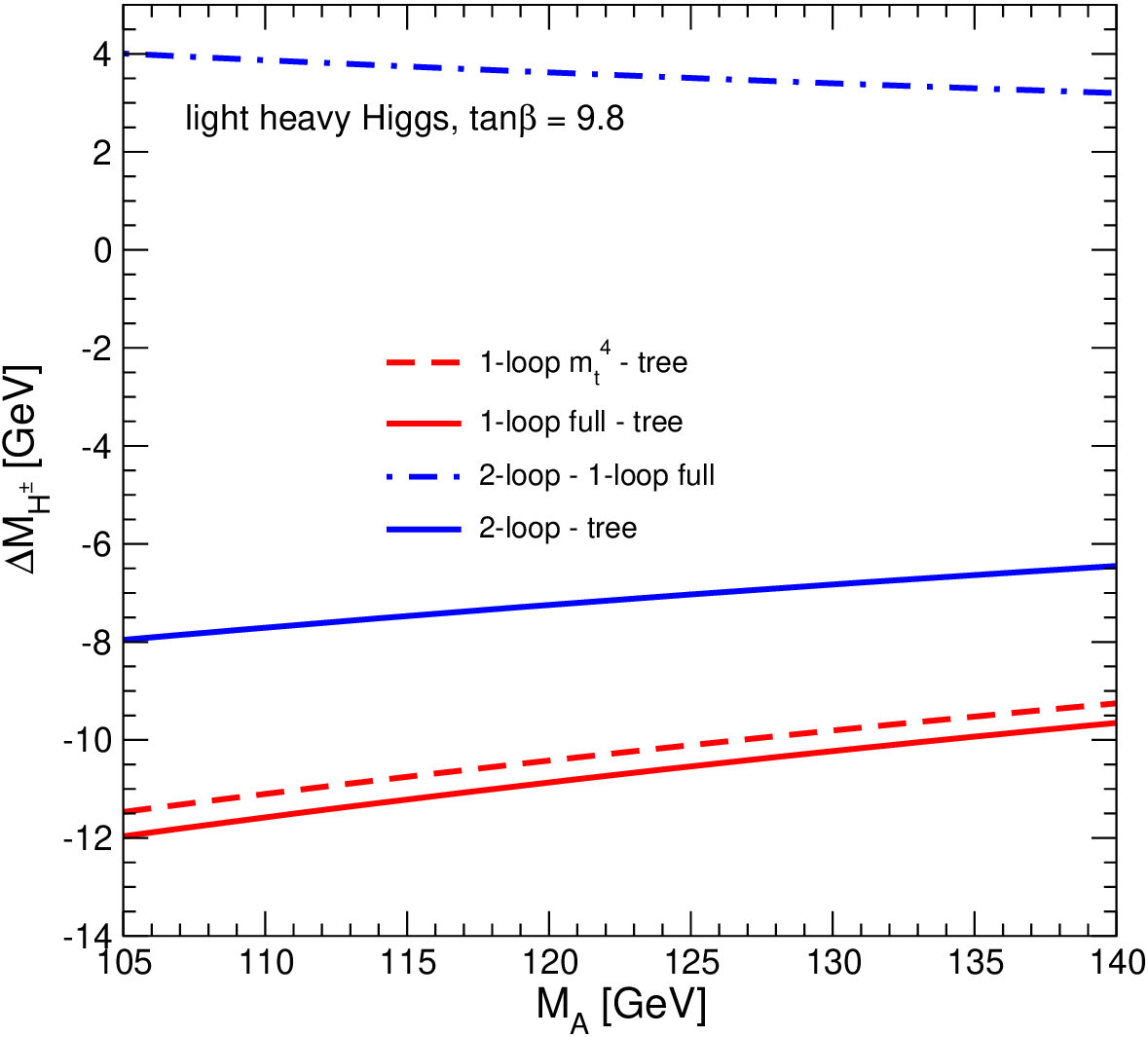}
            \hspace{1em}
            \includegraphics[width=0.5\linewidth]{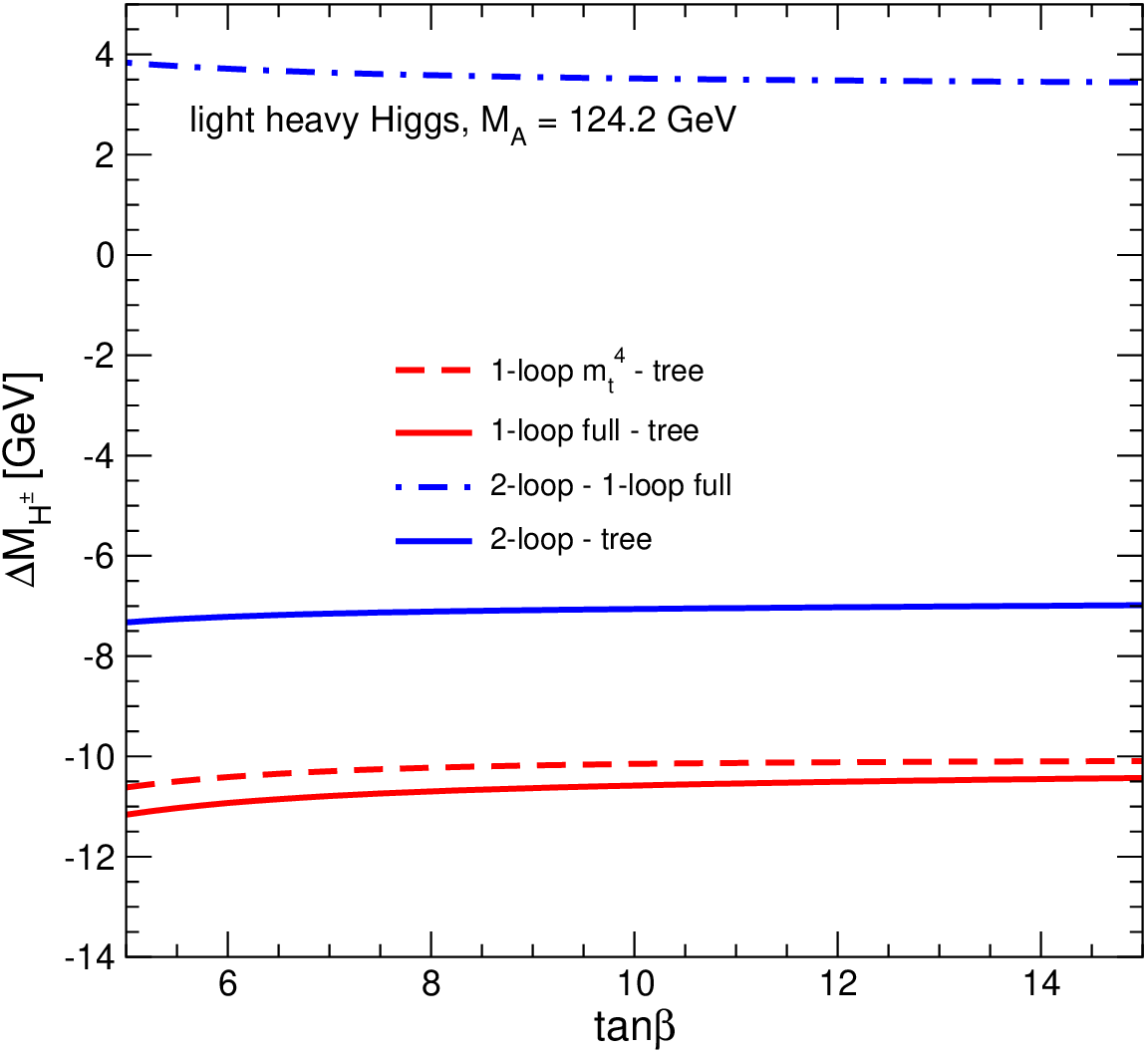}}
\fcaption{$\De\MHp := \MHp - \mHp$ 
  and $\MHp^{2-{\rm loop}} - \MHp^{1-{\rm loop}}$ are
  shown in the ``light heavy Higgs'' scenario, as a function of $\MA$
  with $\tb = 9.8$ (left) and as a function of $\tb$ for $\MA = 124.2 \gev$
  (right). 
  $\MHp$ is evaluated at
  the two-loop level (blue/dark gray solid).
  Also shown are the corresponding one-loop results of
  \order{\mt^4/\MW^2} (dashed), the full one-loop
  corrections (red/light gray solid) and the difference between the two-loop
  and the full one-loop result (dot-dashed).
}
\label{fig:DeMHp2L_lhH}
\end{figure}

Finally, in \reffi{fig:DeMHp2L_lhH}, we analyze the two-loop corrections
to $\MHp$ in the ``light heavy Higgs'' scenario, in which the {\em heavy}
$\cp$-even Higgs boson is interpreted as the newly discovered particle
at $\sim \MHexp \gev$~\cite{hifi}. We show the results as a function of
$\MA$ (left) and $\tb$ (right) with the other parameters fixed as in
\refeq{lhH}. 
This scenario is characterized by a very rich phenomenology,
since all five Higgs states in this case are rather light. Such a
scenario can be probed at the LHC via searches for the heavier neutral
Higgs bosons, $H$ and $A$, but also searches for a light charged
Higgs boson that is produced in top quark decays are of particular
relevance in this case.
As can be seen in \reffi{fig:DeMHp2L_lhH} the $\mt^4/\MW^2$
corrections are an excellent approximation for the full one-loop result
in the parameter space analyzed. 
The one-loop corrections are found to be large and negative in this
case, while the two-loop corrections are positive and at the level of 
$3.5 \gev$ to $4 \gev$, amounting to about 30\% of the one-loop
corrections. Clearly, a thorough treatment of the higher-order
contributions will be important for exploring the charged Higgs boson
phenomenology in such a scenario.


\section{Conclusions}
\label{sec:conclusions}

We have presented a detailed analysis of the 
prediction for the charged Higgs boson mass, $\MHp$, within the MSSM,
on the basis of a complete one-loop calculation, 
and incorporating the two-loop contributions of \order{\alt\als}.

We find relatively large mass shifts at the one-loop level. 
In particular, we have analyzed the dependence of $\MHp$ on the
trilinear couplings $\At$ and $\Ab$. 
For the case $\At = \Ab$, which is assumed in the
$\mhmax$ benchmark scenario, corrections to $\MHp$ of several GeV are found.
The opposite case, $\At = -\Ab$, can yield much larger shifts
exceeding $\De\MHp = 10 \gev$ for large $|\At|$.
In general, the full
one-loop corrections are negative for small $\tb$ and positive
for large $\tb$ in the $\mhmax$ benchmark scenario. 

Pronounced effects on  $\MHp$ in the region of large $\tb$ originate 
from the standard QCD corrections to the bottom Yukawa coupling,
formally a contribution beyond one-loop order. 
Similarly important are the shifts 
from the inclusion of $\db$ effects, leading to a strong
dependence of $\MHp$ on the size and the sign of $\mu$. 
The contributions from the (s)fermion sector beyond
$t/\Stop/b/\Sbot$ are sizable for $\msusy \gsim 1000 \gev$ and can
exceed $\sim 2 \gev$.

The new two-loop corrections of \order{\alt\als} in most 
of the considered cases
are of opposite sign to the one-loop corrections. 
The induced
shifts in $\MHp$ can be of several GeV for small $\MA$ and
$\tb$ and large values of $|\mu|$, and  are thus of a size
that may be probed at the LHC and the LC.
The set of two-loop contributions considered here 
are expected to be particularly relevant for those MSSM parameter regions
where the $\mt^4/\MW^2$ terms yield a good
approximation to the full one-loop result, i.e.\ in particular 
for relatively low values of $\tb$.
For the general case, a more comprehensive higher-order
calculation would be required.

In particular, we analyzed the size of the \order{\alt\als}
corrections in the ``light heavy Higgs'' scenario, in which the 
{\em heavy} $\cp$-even Higgs boson is interpreted as the newly
discovered particle at $\sim \MHexp \gev$. In this scenario 
all MSSM Higgs bosons are relatively light, and there are 
interesting prospects for charged Higgs searches in top quark decays. 
The $\mt^4/\MW^2$ corrections yield
an excellent approximation of the full one-loop result in this
scenario. The genuine two-loop corrections are found to be up to $4 \gev$, 
and thus are important for investigating charged Higgs phenomenology.

Our results for the charged Higgs-boson mass
are implemented into the public Fortran
code \fh. The code also contains the evaluation of the charged Higgs-boson
decays and the main charged Higgs-boson production channels at the LHC.
The code can be obtained from {\tt www.feynhiggs.de}\,.


\subsection*{Acknowledgements}

Work supported in part by the European Community's Marie-Curie Research
Training Network under contract MRTN-CT-2006-035505
`Tools and Precision Calculations for Physics Discoveries at Colliders'.
The work of S.H.\ was supported in part by CICYT 
(grant FPA 2010--22163-C02-01) and by the Spanish MICINN's 
Consolider-Ingenio 2010 Program under grant MultiDark CSD2009-00064.
This work has been supported by the Collaborative Research Center
SFB676 of the DFG, ``Particles, Strings, and the Early Universe''.




\end{document}

%% file: fdSEc.tex
\begin{feynartspicture}(1600,1600)(5,4)
\footnotesize


\FADiagram{}
\FAProp(0.,10.)(10.,10.)(0.,){/ScalarDash}{1}
\FAProp(20.,10.)(10.,10.)(0.,){/ScalarDash}{-1}
\FAProp(10.,10.)(10.,10.)(10.,15.5){/ScalarDash}{-1}
\FALabel(10.,16.57)[b]{$\tilde{\nu}_l$}
\FAVert(10.,10.){0}

\FADiagram{}
\FAProp(0.,10.)(10.,10.)(0.,){/ScalarDash}{1}
\FAProp(20.,10.)(10.,10.)(0.,){/ScalarDash}{-1}
\FAProp(10.,10.)(10.,10.)(10.,15.5){/ScalarDash}{-1}
\FALabel(10.,16.57)[b]{$\{\tilde{l},\tilde{u},\tilde{d}\}_1,\{\tilde{l},\tilde{u},\tilde{d}\}_2$}
\FAVert(10.,10.){0}

\FADiagram{}
\FAProp(0.,10.)(6.,10.)(0.,){/ScalarDash}{1}
\FAProp(20.,10.)(14.,10.)(0.,){/ScalarDash}{-1}
\FAProp(6.,10.)(14.,10.)(0.8,){/Straight}{-1}
\FALabel(10.,5.73)[t]{$\nu_l$}
\FAProp(6.,10.)(14.,10.)(-0.8,){/Straight}{1}
\FALabel(10.,14.27)[b]{$l$}
\FAVert(6.,10.){0}
\FAVert(14.,10.){0}

\FADiagram{}
\FAProp(0.,10.)(6.,10.)(0.,){/ScalarDash}{1}
\FAProp(20.,10.)(14.,10.)(0.,){/ScalarDash}{-1}
\FAProp(6.,10.)(14.,10.)(0.8,){/Straight}{-1}
\FALabel(10.,5.73)[t]{$d$}
\FAProp(6.,10.)(14.,10.)(-0.8,){/Straight}{1}
\FALabel(10.,14.27)[b]{$u$}
\FAVert(6.,10.){0}
\FAVert(14.,10.){0}

\FADiagram{}
\FAProp(0.,10.)(6.,10.)(0.,){/ScalarDash}{1}
\FAProp(20.,10.)(14.,10.)(0.,){/ScalarDash}{-1}
\FAProp(6.,10.)(14.,10.)(0.8,){/ScalarDash}{-1}
\FALabel(10.,5.73)[t]{$\tilde{\nu}_l$}
\FAProp(6.,10.)(14.,10.)(-0.8,){/ScalarDash}{1}
\FALabel(10.,14.27)[b]{$\tilde{l}_1,\tilde{l}_2$}
\FAVert(6.,10.){0}
\FAVert(14.,10.){0}

\FADiagram{}
\FAProp(0.,10.)(6.,10.)(0.,){/ScalarDash}{1}
\FAProp(20.,10.)(14.,10.)(0.,){/ScalarDash}{-1}
\FAProp(6.,10.)(14.,10.)(0.8,){/ScalarDash}{-1}
\FALabel(10.,5.73)[t]{$\tilde{u}_1,\tilde{u}_2$}
\FAProp(6.,10.)(14.,10.)(-0.8,){/ScalarDash}{1}
\FALabel(10.,14.27)[b]{$\tilde{d}_1,\tilde{d}_2$}
\FAVert(6.,10.){0}
\FAVert(14.,10.){0}

\FADiagram{}
\FAProp(0.,10.)(10.,10.)(0.,){/ScalarDash}{1}
\FAProp(20.,10.)(10.,10.)(0.,){/ScalarDash}{-1}
\FAProp(10.,10.)(10.,10.)(10.,15.5){/ScalarDash}{0}
\FALabel(10.,16.32)[b]{$h,H,A,G$}
\FAVert(10.,10.){0}

\FADiagram{}
\FAProp(0.,10.)(10.,10.)(0.,){/ScalarDash}{1}
\FAProp(20.,10.)(10.,10.)(0.,){/ScalarDash}{-1}
\FAProp(10.,10.)(10.,10.)(10.,15.5){/ScalarDash}{-1}
\FALabel(10.,16.57)[b]{$H^\pm,G^\pm$}
\FAVert(10.,10.){0}

\FADiagram{}
\FAProp(0.,10.)(10.,10.)(0.,){/ScalarDash}{1}
\FAProp(20.,10.)(10.,10.)(0.,){/ScalarDash}{-1}
\FAProp(10.,10.)(10.,10.)(10.,15.5){/Sine}{0}
\FALabel(10.,16.57)[b]{$Z$}
\FAVert(10.,10.){0}

\FADiagram{}
\FAProp(0.,10.)(10.,10.)(0.,){/ScalarDash}{1}
\FAProp(20.,10.)(10.,10.)(0.,){/ScalarDash}{-1}
\FAProp(10.,10.)(10.,10.)(10.,15.5){/Sine}{-1}
\FALabel(10.,16.57)[b]{$W^\pm$}
\FAVert(10.,10.){0}

\FADiagram{}
\FAProp(0.,10.)(6.,10.)(0.,){/ScalarDash}{1}
\FAProp(20.,10.)(14.,10.)(0.,){/ScalarDash}{-1}
\FAProp(6.,10.)(14.,10.)(0.8,){/Straight}{0}
\FALabel(10.,5.98)[t]{$\tilde{\chi}^0_1,\tilde{\chi}^0_2,\tilde{\chi}^0_3,\tilde{\chi}^0_4$}
\FAProp(6.,10.)(14.,10.)(-0.8,){/Straight}{1}
\FALabel(10.,14.02)[b]{$\tilde{\chi}^\pm_1,\tilde{\chi}^\pm_2$}
\FAVert(6.,10.){0}
\FAVert(14.,10.){0}


\FADiagram{}
\FAProp(0.,10.)(6.,10.)(0.,){/ScalarDash}{1}
\FAProp(20.,10.)(14.,10.)(0.,){/ScalarDash}{-1}
\FAProp(6.,10.)(14.,10.)(0.8,){/ScalarDash}{0}
\FALabel(10.,5.73)[t]{$h,H,A$}
\FAProp(6.,10.)(14.,10.)(-0.8,){/ScalarDash}{1}
\FALabel(10.,14.27)[b]{$H^\pm$}
\FAVert(6.,10.){0}
\FAVert(14.,10.){0}


\FADiagram{}
\FAProp(0.,10.)(6.,10.)(0.,){/ScalarDash}{1}
\FAProp(20.,10.)(14.,10.)(0.,){/ScalarDash}{-1}
\FAProp(6.,10.)(14.,10.)(0.8,){/ScalarDash}{0}
\FALabel(10.,5.73)[t]{$h,H,A,G$}
\FAProp(6.,10.)(14.,10.)(-0.8,){/ScalarDash}{1}
\FALabel(10.,14.27)[b]{$G^\pm$}
\FAVert(6.,10.){0}
\FAVert(14.,10.){0}



\FADiagram{}
\FAProp(0.,10.)(6.,10.)(0.,){/ScalarDash}{1}
\FAProp(20.,10.)(14.,10.)(0.,){/ScalarDash}{-1}
\FAProp(6.,10.)(14.,10.)(0.8,){/ScalarDash}{1}
\FALabel(10.,5.98)[t]{$H^\pm,G^\pm$}
\FAProp(6.,10.)(14.,10.)(-0.8,){/Sine}{0}
\FALabel(10.,14.27)[b]{$\gamma,Z$}
\FAVert(6.,10.){0}
\FAVert(14.,10.){0}


\FADiagram{}
\FAProp(0.,10.)(6.,10.)(0.,){/ScalarDash}{1}
\FAProp(20.,10.)(14.,10.)(0.,){/ScalarDash}{-1}
\FAProp(6.,10.)(14.,10.)(0.8,){/ScalarDash}{0}
\FALabel(10.,5.98)[t]{$h,H,A,G$}
\FAProp(6.,10.)(14.,10.)(-0.8,){/Sine}{1}
\FALabel(10.,14.27)[b]{$W^\pm$}
\FAVert(6.,10.){0}
\FAVert(14.,10.){0}



\FADiagram{}
\FAProp(0.,10.)(6.,10.)(0.,){/ScalarDash}{1}
\FAProp(20.,10.)(14.,10.)(0.,){/ScalarDash}{-1}
\FAProp(6.,10.)(14.,10.)(0.8,){/Sine}{0}
\FALabel(10.,5.73)[t]{$\gamma,Z$}
\FAProp(6.,10.)(14.,10.)(-0.8,){/Sine}{1}
\FALabel(10.,14.27)[b]{$W^\pm$}
\FAVert(6.,10.){0}
\FAVert(14.,10.){0}

\FADiagram{}
\FAProp(0.,10.)(6.,10.)(0.,){/ScalarDash}{1}
\FAProp(20.,10.)(14.,10.)(0.,){/ScalarDash}{-1}
\FAProp(6.,10.)(14.,10.)(0.8,){/GhostDash}{-1}
\FALabel(10.,5.73)[t]{$u_Z$}
\FAProp(6.,10.)(14.,10.)(-0.8,){/GhostDash}{1}
\FALabel(10.,14.27)[b]{$u^-$}
\FAVert(6.,10.){0}
\FAVert(14.,10.){0}

\FADiagram{}
\FAProp(0.,10.)(6.,10.)(0.,){/ScalarDash}{1}
\FAProp(20.,10.)(14.,10.)(0.,){/ScalarDash}{-1}
\FAProp(6.,10.)(14.,10.)(0.8,){/GhostDash}{1}
\FALabel(10.,5.73)[t]{$u_Z$}
\FAProp(6.,10.)(14.,10.)(-0.8,){/GhostDash}{-1}
\FALabel(10.,14.27)[b]{$u^+$}
\FAVert(6.,10.){0}
\FAVert(14.,10.){0}

\FADiagram{}
\FAProp(0.,10.)(6.,10.)(0.,){/ScalarDash}{1}
\FAProp(20.,10.)(14.,10.)(0.,){/ScalarDash}{-1}
\FAProp(6.,10.)(14.,10.)(0.8,){/GhostDash}{-1}
\FALabel(10.,5.73)[t]{$u_\gamma$}
\FAProp(6.,10.)(14.,10.)(-0.8,){/GhostDash}{1}
\FALabel(10.,14.27)[b]{$u^-$}
\FAVert(6.,10.){0}
\FAVert(14.,10.){0}

\FADiagram{}
\FAProp(0.,10.)(6.,10.)(0.,){/ScalarDash}{1}
\FAProp(20.,10.)(14.,10.)(0.,){/ScalarDash}{-1}
\FAProp(6.,10.)(14.,10.)(0.8,){/GhostDash}{1}
\FALabel(10.,5.73)[t]{$u_\gamma$}
\FAProp(6.,10.)(14.,10.)(-0.8,){/GhostDash}{-1}
\FALabel(10.,14.27)[b]{$u^+$}
\FAVert(6.,10.){0}
\FAVert(14.,10.){0}

\end{feynartspicture}

%% file: fdSEn.tex
\begin{feynartspicture}(1600,2000)(5,6)
\footnotesize

\FADiagram{}
\FAProp(0.,10.)(10.,10.)(0.,){/ScalarDash}{0}
\FAProp(20.,10.)(10.,10.)(0.,){/ScalarDash}{0}
\FAProp(10.,10.)(10.,10.)(10.,15.5){/ScalarDash}{-1}
\FALabel(10.,16.57)[b]{$\tilde{\nu}_{\{e,\mu,\tau\}}$}
\FAVert(10.,10.){0}

\FADiagram{}
\FAProp(0.,10.)(10.,10.)(0.,){/ScalarDash}{0}
\FAProp(20.,10.)(10.,10.)(0.,){/ScalarDash}{0}
\FAProp(10.,10.)(10.,10.)(10.,15.5){/ScalarDash}{-1}
\FALabel(10.,16.57)[b]{$\tilde{f}_1,\tilde{f}_2$}
\FAVert(10.,10.){0}

\FADiagram{}
\FAProp(0.,10.)(6.,10.)(0.,){/ScalarDash}{0}
\FAProp(20.,10.)(14.,10.)(0.,){/ScalarDash}{0}
\FAProp(6.,10.)(14.,10.)(0.8,){/Straight}{-1}
\FALabel(10.,5.73)[t]{$\nu_{\{e,\mu,\tau\}}$}
\FAProp(6.,10.)(14.,10.)(-0.8,){/Straight}{1}
\FALabel(10.,14.27)[b]{$\nu_{\{e,\mu,\tau\}}$}
\FAVert(6.,10.){0}
\FAVert(14.,10.){0}

\FADiagram{}
\FAProp(0.,10.)(6.,10.)(0.,){/ScalarDash}{0}
\FAProp(20.,10.)(14.,10.)(0.,){/ScalarDash}{0}
\FAProp(6.,10.)(14.,10.)(0.8,){/Straight}{-1}
\FALabel(10.,5.73)[t]{$f$}
\FAProp(6.,10.)(14.,10.)(-0.8,){/Straight}{1}
\FALabel(10.,14.27)[b]{$f$}
\FAVert(6.,10.){0}
\FAVert(14.,10.){0}

\FADiagram{}
\FAProp(0.,10.)(6.,10.)(0.,){/ScalarDash}{0}
\FAProp(20.,10.)(14.,10.)(0.,){/ScalarDash}{0}
\FAProp(6.,10.)(14.,10.)(0.8,){/ScalarDash}{-1}
\FALabel(10.,5.73)[t]{$\tilde{\nu}_{\{e,\mu,\tau\}}$}
\FAProp(6.,10.)(14.,10.)(-0.8,){/ScalarDash}{1}
\FALabel(10.,14.27)[b]{$\tilde{\nu}_{\{e,\mu,\tau\}}$}
\FAVert(6.,10.){0}
\FAVert(14.,10.){0}

\FADiagram{}
\FAProp(0.,10.)(6.,10.)(0.,){/ScalarDash}{0}
\FAProp(20.,10.)(14.,10.)(0.,){/ScalarDash}{0}
\FAProp(6.,10.)(14.,10.)(0.8,){/ScalarDash}{-1}
\FALabel(10.,5.73)[t]{$\tilde{f}_1,\tilde{f}_2$}
\FAProp(6.,10.)(14.,10.)(-0.8,){/ScalarDash}{1}
\FALabel(10.,14.27)[b]{$\tilde{f}_1,\tilde{f}_2$}
\FAVert(6.,10.){0}
\FAVert(14.,10.){0}

\FADiagram{}
\FAProp(0.,10.)(10.,10.)(0.,){/ScalarDash}{0}
\FAProp(20.,10.)(10.,10.)(0.,){/ScalarDash}{0}
\FAProp(10.,10.)(10.,10.)(10.,15.5){/ScalarDash}{-1}
\FALabel(10.,16.57)[b]{$H^\pm,G^\pm$}
\FAVert(10.,10.){0}

\FADiagram{}
\FAProp(0.,10.)(10.,10.)(0.,){/ScalarDash}{0}
\FAProp(20.,10.)(10.,10.)(0.,){/ScalarDash}{0}
\FAProp(10.,10.)(10.,10.)(10.,15.5){/Sine}{-1}
\FALabel(10.,16.57)[b]{$W^\pm$}
\FAVert(10.,10.){0}

\FADiagram{}
\FAProp(0.,10.)(6.,10.)(0.,){/ScalarDash}{0}
\FAProp(20.,10.)(14.,10.)(0.,){/ScalarDash}{0}
\FAProp(6.,10.)(14.,10.)(0.8,){/Straight}{-1}
\FALabel(10.,5.73)[t]{$\tilde{\chi}^\pm_1,\tilde{\chi}^\pm_2$}
\FAProp(6.,10.)(14.,10.)(-0.8,){/Straight}{1}
\FALabel(10.,14.27)[b]{$\tilde{\chi}^\pm_1,\tilde{\chi}^\pm_2$}
\FAVert(6.,10.){0}
\FAVert(14.,10.){0}

\FADiagram{}
\FAProp(0.,10.)(6.,10.)(0.,){/ScalarDash}{0}
\FAProp(20.,10.)(14.,10.)(0.,){/ScalarDash}{0}
\FAProp(6.,10.)(14.,10.)(0.8,){/ScalarDash}{-1}
\FALabel(10.,5.73)[t]{$H^\pm$}
\FAProp(6.,10.)(14.,10.)(-0.8,){/ScalarDash}{1}
\FALabel(10.,14.27)[b]{$H^\pm$}
\FAVert(6.,10.){0}
\FAVert(14.,10.){0}

\FADiagram{}
\FAProp(0.,10.)(6.,10.)(0.,){/ScalarDash}{0}
\FAProp(20.,10.)(14.,10.)(0.,){/ScalarDash}{0}
\FAProp(6.,10.)(14.,10.)(0.8,){/ScalarDash}{-1}
\FALabel(10.,5.73)[t]{$G^\pm$}
\FAProp(6.,10.)(14.,10.)(-0.8,){/ScalarDash}{1}
\FALabel(10.,14.27)[b]{$H^\pm$}
\FAVert(6.,10.){0}
\FAVert(14.,10.){0}

\FADiagram{}
\FAProp(0.,10.)(6.,10.)(0.,){/ScalarDash}{0}
\FAProp(20.,10.)(14.,10.)(0.,){/ScalarDash}{0}
\FAProp(6.,10.)(14.,10.)(0.8,){/ScalarDash}{1}
\FALabel(10.,5.73)[t]{$G^\pm$}
\FAProp(6.,10.)(14.,10.)(-0.8,){/ScalarDash}{-1}
\FALabel(10.,14.27)[b]{$G^\pm$}
\FAVert(6.,10.){0}
\FAVert(14.,10.){0}


\FADiagram{}
\FAProp(0.,10.)(6.,10.)(0.,){/ScalarDash}{0}
\FAProp(20.,10.)(14.,10.)(0.,){/ScalarDash}{0}
\FAProp(6.,10.)(14.,10.)(0.8,){/GhostDash}{-1}
\FALabel(10.,5.73)[t]{$u^\pm$}
\FAProp(6.,10.)(14.,10.)(-0.8,){/GhostDash}{1}
\FALabel(10.,14.27)[b]{$u^\pm$}
\FAVert(6.,10.){0}
\FAVert(14.,10.){0}

\FADiagram{}
\FAProp(0.,10.)(6.,10.)(0.,){/ScalarDash}{0}
\FAProp(20.,10.)(14.,10.)(0.,){/ScalarDash}{0}
\FAProp(6.,10.)(14.,10.)(0.8,){/Sine}{-1}
\FALabel(10.,5.73)[t]{$W^\pm$}
\FAProp(6.,10.)(14.,10.)(-0.8,){/Sine}{1}
\FALabel(10.,14.27)[b]{$W^\pm$}
\FAVert(6.,10.){0}
\FAVert(14.,10.){0}

\FADiagram{}
\FAProp(0.,10.)(6.,10.)(0.,){/ScalarDash}{0}
\FAProp(20.,10.)(14.,10.)(0.,){/ScalarDash}{0}
\FAProp(6.,10.)(14.,10.)(0.8,){/ScalarDash}{1}
\FALabel(10.,5.73)[t]{$H^\pm$}
\FAProp(6.,10.)(14.,10.)(-0.8,){/Sine}{-1}
\FALabel(10.,14.27)[b]{$W^\pm$}
\FAVert(6.,10.){0}
\FAVert(14.,10.){0}

\FADiagram{}
\FAProp(0.,10.)(6.,10.)(0.,){/ScalarDash}{0}
\FAProp(20.,10.)(14.,10.)(0.,){/ScalarDash}{0}
\FAProp(6.,10.)(14.,10.)(0.8,){/ScalarDash}{1}
\FALabel(10.,5.73)[t]{$G^\pm$}
\FAProp(6.,10.)(14.,10.)(-0.8,){/Sine}{-1}
\FALabel(10.,14.27)[b]{$W^\pm$}
\FAVert(6.,10.){0}
\FAVert(14.,10.){0}

\FADiagram{}
\FAProp(0.,10.)(6.,10.)(0.,){/ScalarDash}{0}
\FAProp(20.,10.)(14.,10.)(0.,){/ScalarDash}{0}
\FAProp(6.,10.)(14.,10.)(0.8,){/ScalarDash}{-1}
\FALabel(10.,5.73)[t]{$H^\pm$}
\FAProp(6.,10.)(14.,10.)(-0.8,){/Sine}{1}
\FALabel(10.,14.27)[b]{$W^\pm$}
\FAVert(6.,10.){0}
\FAVert(14.,10.){0}

\FADiagram{}
\FAProp(0.,10.)(6.,10.)(0.,){/ScalarDash}{0}
\FAProp(20.,10.)(14.,10.)(0.,){/ScalarDash}{0}
\FAProp(6.,10.)(14.,10.)(0.8,){/ScalarDash}{-1}
\FALabel(10.,5.73)[t]{$G^\pm$}
\FAProp(6.,10.)(14.,10.)(-0.8,){/Sine}{1}
\FALabel(10.,14.27)[b]{$W^\pm$}
\FAVert(6.,10.){0}
\FAVert(14.,10.){0}

\FADiagram{}
\FAProp(0.,10.)(10.,10.)(0.,){/ScalarDash}{0}
\FAProp(20.,10.)(10.,10.)(0.,){/ScalarDash}{0}
\FAProp(10.,10.)(10.,10.)(10.,15.5){/ScalarDash}{0}
\FALabel(10.,16.32)[b]{$h,H,A,G$}
\FAVert(10.,10.){0}

\FADiagram{}
\FAProp(0.,10.)(10.,10.)(0.,){/ScalarDash}{0}
\FAProp(20.,10.)(10.,10.)(0.,){/ScalarDash}{0}
\FAProp(10.,10.)(10.,10.)(10.,15.5){/Sine}{0}
\FALabel(10.,16.57)[b]{$Z$}
\FAVert(10.,10.){0}




\FADiagram{}
\FAProp(0.,10.)(6.,10.)(0.,){/ScalarDash}{0}
\FAProp(20.,10.)(14.,10.)(0.,){/ScalarDash}{0}
\FAProp(6.,10.)(14.,10.)(0.8,){/Straight}{0}
\FALabel(10.,5.98)[t]{$\tilde{\chi}^0_1,\tilde{\chi}^0_2,\tilde{\chi}^0_3,\tilde{\chi}^0_4$}
\FAProp(6.,10.)(14.,10.)(-0.8,){/Straight}{0}
\FALabel(10.,14.02)[b]{$\tilde{\chi}^0_1,\tilde{\chi}^0_2,\tilde{\chi}^0_3,\tilde{\chi}^0_4$}
\FAVert(6.,10.){0}
\FAVert(14.,10.){0}

\FADiagram{}
\FAProp(0.,10.)(6.,10.)(0.,){/ScalarDash}{0}
\FAProp(20.,10.)(14.,10.)(0.,){/ScalarDash}{0}
\FAProp(6.,10.)(14.,10.)(0.8,){/ScalarDash}{0}
\FALabel(10.,5.98)[t]{$h,H,A,G$}
\FAProp(6.,10.)(14.,10.)(-0.8,){/ScalarDash}{0}
\FALabel(10.,14.02)[b]{$h,H,A,G$}
\FAVert(6.,10.){0}
\FAVert(14.,10.){0}






\FADiagram{}
\FAProp(0.,10.)(6.,10.)(0.,){/ScalarDash}{0}
\FAProp(20.,10.)(14.,10.)(0.,){/ScalarDash}{0}
\FAProp(6.,10.)(14.,10.)(0.8,){/GhostDash}{-1}
\FALabel(10.,5.73)[t]{$u_Z$}
\FAProp(6.,10.)(14.,10.)(-0.8,){/GhostDash}{1}
\FALabel(10.,14.27)[b]{$u_Z$}
\FAVert(6.,10.){0}
\FAVert(14.,10.){0}

\FADiagram{}
\FAProp(0.,10.)(6.,10.)(0.,){/ScalarDash}{0}
\FAProp(20.,10.)(14.,10.)(0.,){/ScalarDash}{0}
\FAProp(6.,10.)(14.,10.)(0.8,){/Sine}{0}
\FALabel(10.,5.73)[t]{$Z$}
\FAProp(6.,10.)(14.,10.)(-0.8,){/Sine}{0}
\FALabel(10.,14.27)[b]{$Z$}
\FAVert(6.,10.){0}
\FAVert(14.,10.){0}

\FADiagram{}
\FAProp(0.,10.)(6.,10.)(0.,){/ScalarDash}{0}
\FAProp(20.,10.)(14.,10.)(0.,){/ScalarDash}{0}
\FAProp(6.,10.)(14.,10.)(0.8,){/ScalarDash}{0}
\FALabel(10.,5.98)[t]{$h,H,A,G$}
\FAProp(6.,10.)(14.,10.)(-0.8,){/Sine}{0}
\FALabel(10.,14.27)[b]{$Z$}
\FAVert(6.,10.){0}
\FAVert(14.,10.){0}


\end{feynartspicture}